\documentclass{aa}
\usepackage{color}
\usepackage{graphics,epsfig}
\usepackage{graphicx}
\usepackage{rotating}

\begin{document}
%
  {\it \bf Based on observations with ISO, an ESA project with
  instruments funded by ESA Member States (especially the PI
  countries: France, Germany, the Netherlands and the United
  Kingdom) and with the participation of ISAS and NASA.}\\

\color{black}

\title{An ISOCAM survey through gravitationally lensing galaxy clusters
\thanks{Based on observations with ISO, an ESA project with instruments
funded by ESA Member States (especially the PI countries: France,
Germany, the Netherlands and the United Kingdom) with the participation
of ISAS and NASA}  
}
       
\subtitle{I. Source lists and source counts for A370, A2218 and A2390}
\author{L.~Metcalfe \inst{1, 4}
   \and J.-P.~Kneib  \inst{2}
   \and B.~McBreen  \inst{3}
   \and B.~Altieri  \inst{1}
   \and A.~Biviano  \inst{5}
   \and M.~Delaney  \inst{3}
   \and D.~Elbaz    \inst{6}
   \and  M.F.~Kessler\inst{7}
   \and K.~Leech    \inst{4}
   \and K.~Okumura  \inst{6}
   \and S.~Ott      \inst{7}
   \and R.~Perez-Martinez \inst{1, 4}
   \and  C.~Sanchez-Fernandez \inst{1, 4} 
   \and B.~Schulz   \inst{4, 8}
}

\institute{XMM-Newton Science Operations Centre, European Space Agency, Villafranca del Castillo, P.O. Box 50727, 28080 Madrid, Spain.
      \and Observatoire Midi-Pyrenees, 14 Av. E. Belin, 31400 Toulouse, France.
      \and Physics Department, University College Dublin, Stillorgan Road, Dublin 4, Ireland.
      \and ISO Data Centre, European Space Agency, Villafranca del Castillo, P.O. Box 50727, 28080 Madrid, Spain.
      \and INAF/Osservatorio Astronomico di Trieste, via G.B. Tiepolo 11, 34131, Trieste, Italy
      \and DSM/DAPNIA Service d'Astrophysique, CEA - Saclay, Bat. 709, Orme des Merisiers, 91191 Gif-sur-Yvette CEDEX, France.
      \and Science Operations and Data Systems Division of ESA, ESTEC, Keplerlaan 1, 2200 AG Noordwijk, The Netherlands.
      \and Infrared Processing and Analysis Center, California Institute of Technology, Pasadena, CA  91125, U.S.A.
}

\offprints{lmetcalf@xmm.vilspa.esa.es}
\date{Received January 07, 2003; accepted April 25, 2003.} 
 
\abstract {
ESA's Infrared Space Observatory (ISO) was used to perform a deep survey
with ISOCAM through three massive gravitationally lensing clusters  of
galaxies.  The total area surveyed depends on source flux, with nearly
seventy square arcminutes covered for the brighter flux levels in maps 
centred on the three clusters Abell 370, Abell 2218 and Abell 2390.  We
present maps and photometry at 6.7\,$\mu$m (hereafter 7\,$\mu$m) and 
14.3\,$\mu$m (hereafter 15\,$\mu$m), showing a total of 145 mid-infrared 
sources and the associated source counts. At 15\,$\mu$m these counts reach 
the faintest level yet recorded.  Almost all of the sources have been
confirmed on more than one infrared map and all are identified with
counterparts in the optical or near-infrared. Detailed models of the
three clusters have been used to correct for the effects of
gravitational lensing on the background source population. Lensing by
the clusters increases the sensitivity of the survey, and
the weakest sources have lensing corrected fluxes of 5 and 18\,$\mu$Jy
at 7 and 15\,$\mu$m, respectively.  Roughly 70\% of the
15\,$\mu$m sources are lensed  background galaxies. Of sources
detected only at 7\,$\mu$m, 95\% are cluster galaxies for this
sample. Of 
fifteen SCUBA sources within the mapped regions of the three clusters 
seven were detected at 15 microns.  The redshifts for five of these
sources lie in the range 0.23 to 2.8, with a median value of 0.9.
\\ 
Flux selected subsets of the field sources above the 80\% and 50\%
completeness limits were used to derive source counts to a lensing
corrected sensitivity level of 30\,$\mu$Jy at 15\,$\mu$m, and
14\,$\mu$Jy at 7\,$\mu$m.   The source counts, corrected for the
effects of completeness, contamination by cluster sources and lensing,
confirm and extend earlier findings of an excess by a factor of ten in
the 15\,$\mu$m population with respect to source models with no
evolution.  The observed mid-infrared field sources occur mostly at
redshifts between 0.4 and 1.5.
\\
For the counts at 7\,$\mu$m, integrating in the range 14\,$\mu$Jy to 
460\,$\mu$Jy, we resolve (0.49$\pm$0.2)$\times$10$^{-9}$\,W\,m$^{-2}$\,sr$^{-1}$ of
the infrared background light into discrete sources. At 15\,$\mu$m we
include the counts from other extensive ISOCAM surveys to integrate
over the range 30$\mu$Jy to 50 mJy, reaching two to three times
deeper than
the unlensed surveys to resolve 
(2.7$\pm$0.62)$\times$10$^{-9}$\,W\,m$^{-2}$\,sr$^{-1}$ of the
infrared background light. These values correspond to 10\% and 55\%,
respectively, of the upper limit to the infrared background light, derived 
from photon-photon pair production of the high energy gamma rays from
BL-Lac sources on the infrared background photons.  However, the recent
detections of TeV gamma rays from the z=0.129 
BL Lac H1426+428 suggest that the value for the 15\,$\mu$m background 
reported here is already sufficient to imply substantial absorption of TeV 
gamma rays from that source.
}


\titlerunning{An ISOCAM survey through lensing clusters}
\authorrunning{Metcalfe et al.}
\maketitle 
 
%

\section{Introduction} 

\subsection{General introduction and background}
\label{sec:lens_hist}

Following the identification of a double quasar as the first recognised
cosmological gravitational lens (Young et al.\,1980), the theoretical and 
observational exploitation of the lensing phenomenon developed rapidly.
In 1986 luminous giant arcs were discovered in the fields of some
galaxy clusters by Lynds \& Petrosian (1986 \& 1989), and independently by
Soucail et al.\,(1987a \& b). These were soon recognised to be Einstein rings
(Paczyn\'ski 1987), this proposition being thereafter quickly confirmed
spectroscopically (Soucail et al. 1988).

Since then observations of cluster-lenses have been extended to 
wavelengths other than the visible (e.g.\ Smail et al. 1997 \& 2002, Cowie et al. 2002
\& Blain et al.\,1999 for the  sub-millimetre region; Bautz et al.\,2000 -  
for the X-ray).  At the same time models describing lensing have been well
understood ({\it e.g.} Kneib et al.\,1996; B\'ezecourt et al.\,1999; Smith et 
al.\,2001) to the point where source counts made through a cluster lens can be
corrected to yield the counts and source fluxes which would be found in
the absence of the lens.

\begin{figure*}[!ht]
\centering
\includegraphics[angle=270,width=\textwidth,bb=80 10 480 775]{3463_f1_c.ps} 
\caption[]{(a) shows a regular rectangular pattern laid down over the ISO contour
map of the cluster A2218, itself overlaid on an optical map of the cluster. This defines a regular region
of sky, as observed in the direction of the lensing cluster.  The
distorted version of the pattern shown in (b) demonstrates how a
regular region of sky, as observed, projects back through the
gravitational lens into smaller regions in the source plane. The background sky has been
amplified and the transformation  of area within each contour of the
grid is proportional to the amplification of the lens over the 
corresponding region of sky. (See also Figs.\ref{a2218lw2}
and \ref{a2218lw3}). 
}
\label{action1}
\end{figure*}

In recent years, observations through massive cluster lenses
with well constrained mass models have played an important role in
extending sub-millimetre (submm) observations to the deepest levels
(Smail et al. 1997 \& 2002; Cowie et al. 2002; Ivison et al.\,2000; Blain et al.\,1999).  
Sufficient submm sources have been resolved to account for a diffuse 
850\,$\mu$m background comparable to that measured by COBE (Fixsen et al.\,1998). 
Blain et al.\,(2000) interpret the submm sources to be a population of distant,
dusty galaxies emitting in the submm waveband, missing from optical
surveys, and tracing dust obscured star formation activity at high
redshift (z\,$\la$\,5). The total energy emitted is five times greater
than inferred from rest-frame UV observations. They discuss the
difficulty of distinguishing between star formation and recycling of
AGN emission, as mechanisms for powering the submm emission.   This
couples to a long-standing question concerning the origin of the
diffuse hard-X-ray background.  The spectrum of this background in the
1 to 100 keV range, peaking  around 30 keV 
(Fig.1 of Wilman et al. 2000b), is too flat to be traced to any known population of
extragalactic sources. However, it could be synthesised if a component
is attributed to highly obscured type 2 AGN (Setti \& Woltjer 1989).

Bautz et al.\,(2000) reported Chandra X-ray observations
through the  lensing cluster Abell 370, with detection of hard X-ray 
counterparts for the two prominent SCUBA sources. They concluded that
their results are consistent with 20\% of submm sources exhibiting
significant AGN emission.  

Meanwhile, deep surveys performed in fields relatively free of
foreground source contamination have resolved into discrete background
sources a significant fraction of the diffuse mid-infrared radiation background.
Elbaz et al. (1999 \& 2002),  Gruppioni et al. (2002), Lari et al.
(2001) and Serjeant et al. (2000), using the mid-infrared camera, ISOCAM
(Cesarsky et al.\,1996), on board ESA's Infrared Space Observatory (ISO) 
mission (Kessler et al.\,1996), have reported the results of deep
15\,$\mu$m field surveys in the Hubble Deep Fields North and
South, in the Marano Field and in the Lockman Hole. 

The mid-infrared-luminous galaxies observed have been interpreted (Aussel et
al.\,1999b; Elbaz et al.\,1999 \& 2002; Serjeant et al. 2000) as a population 
of dust-enshrouded star forming galaxies. Within the detection limits
the 15\,$\mu$m population has a median redshift of about 0.8.

Mushotsky et al.\,(2000) reported the resolution, by
Chandra, of at least 75\% of the 2 to 10 keV X-ray background into
discrete sources. Hasinger  et al. (2001) also resolve the hard X-ray
background and in addition benefit from the higher energy limit of
XMM-Newton to study the hard X-ray sources. More recently, Rosati et
al. (2002) further extended the Chandra number counts to resolve up to
90\% of the X-ray background, but positing a non-negligible population
of faint hard  sources still to be discovered at energies not well
probed by Chandra.

Franceschini et al.\,(2002) and Fadda et al.\,(2002) have  investigated
the relationship between the mid-infrared population seen by ISO and
the Chandra and XMM-Newton X-ray sources responsible for the X-ray
background, finding an upper limit of about 20\% to the contribution 
of AGN to the mid-infrared population. 

Against the background of ongoing efforts to achieve the deepest
possible  surveys on the well established deep fields, and considering
the emerging potential of lensing work to allow amplified surveys to be
performed through gravitationally lensing galaxy clusters, we initiated
a gravitational lensing deep-survey programme  over a decade ago as
part of the ``Central Programme'' of the ISO mission. This has extended  the
exploitation of the gravitational lensing phenomenon to the
mid-infrared. The programme has achieved deep and ultra-deep
imaging through a sample of gravitationally lensing clusters of
galaxies, benefiting from the lensing to search for background objects
to a depth not otherwise achievable (Metcalfe et al.\,2001; Altieri et
al.\,1999, Barvainis et al.\,1999).  In this paper we report the
results which have been obtained in the mid-infrared providing, not
only deep counts of the lensed sources,  but extensive mid-infrared
photometry of the lensing clusters themselves. In future work we will
compare the mid-infrared results with available archival data at other
wavelengths.

In Sec.\ref{sec:lens} some of the benefits of  observing through a
gravitational lens are discussed with specific reference to the
properties  of this survey.  Section~\ref{sec:obsred} describes the
observations performed. Section~\ref{sec:analyse} describes the several
steps of the data reduction. This includes the use of Monte-Carlo
simulations to calibrate the faint-source data and assign completeness
limits to the survey. The method employed to correct the results for
the effects of the lensing amplification of the cluster is described.

Section~\ref{sec:results_ims} presents maps of the fields showing  IR
contours at 7\,$\mu$m and 15\,$\mu$m overlaid on optical images, and
also tables of the  7\,$\mu$m and 15\,$\mu$m sources detected, giving
measured coordinates and flux densities. 

Section~\ref{sec:counts} gives 7\,$\mu$m and 15\,$\mu$m source counts
based upon a subset of the sources listed in the tables of
Sec.\ref{sec:results_ims}, and includes an estimate of the
contribution of the source populations to the Infrared Background Light
(IBL) at the above wavelengths.

The results are discussed in Sec.\ref{sec:discussion}, and the conclusions 
stated in Sec.\ref{sec:conclude}



\section{The benefits of observing through a gravitational lens}
\label{sec:lens}

Lensing amplifies sources and, for a given apparent flux limit,
suppresses confusion due to the apparent surface area dilation.


Fig.~\ref{action1} describes the action of a lensing cluster.
Fig.~\ref{action1}a shows a regular rectangular pattern laid down over
the field of the lensing cluster Abell 2218. Fig.~\ref{action1}b shows
how that region of sky is mapped by the lens onto a plane on the
background sky at a redshift of 1, typical of our sources. The
expansion of apparent surface area in passing from the source plane (b)
to the observed sky (a) is very significant near the centre of the
lens, and commensurate flux amplification and reduction of source
confusion at a given flux follow. The amplification in the inner regions exceeds 5 over
a significant area (typically $\sim$\,1\,arcmin$^2$, depending on the
cluster),  and factors of more than 1.5 to 2 apply over much of the
field illustrated.

\begin{figure*}[!ht]
\centering
\includegraphics[angle=270,width=\textwidth,bb=20 25 520 740]{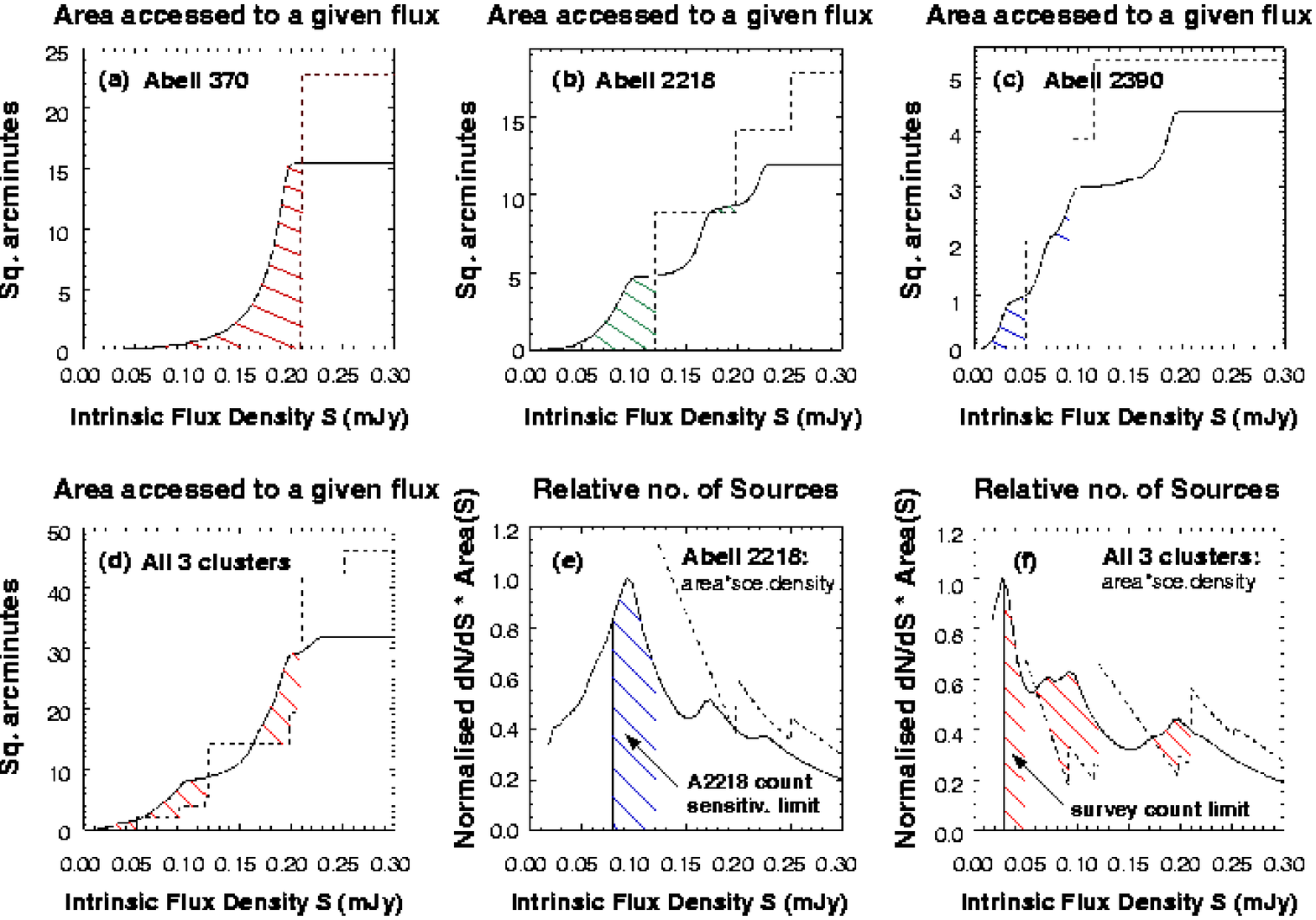}
\caption[]{Increased sensitivity due to lensing:
In (a), (b) and (c) the dashed line defines, per cluster field, the
area of sky accessible to the survey as a function
of flux in the absence of lensing. The profile is stepped because the
rastering measurement performed on the sky covered different regions
to different depth. The solid curve in each case shows how the area of
sky covered depends on the limiting 15\,$\mu$m flux after lensing is taken into
account. (d) shows the same parameters for the composite survey
which results from combining observations on the three clusters. In
(e) and (f) the measured distribution of sources with flux density (dN/dS) has been
folded with the previously illustrated flux-dependent accessible area
in such a way that the resulting curves - (e) for the example of A2218 and (f) for
the three-cluster survey - show the relative number of sources accessible
to a lensing survey at each flux level. It is seen that a significant 
population of sources below the unlensed sensitivity limit (the dashed line) 
becomes accessible to the measurement. In each of (e) and (f) a solid vertical line
delineates the 50\% completeness limit for source counts.  The dashed (unlensed)
case also refers to the 50\% completeness limit in the absence of lensing. The peaks in 
the source-density plots are due to the discontinuities in the sensitivity levels in the
``accessible-areas'' plots, arising from the raster measuring strategy used.}
\label{how_deep1}
\end{figure*}

Fig.~\ref{how_deep1} illustrates an important benefit of
observing through a gravitational lens.  As described in the caption, it compares 
the surface areas of sky covered to a given limiting flux, with and without the 
action of the lensing cluster. The dashed lines in the figure refer to the 
unlensed case. The solid lines in Fig.~\ref{how_deep1}.a through Fig.~\ref{how_deep1}.d 
trace the relationship between achieved sensitivity and
associated area surveyed, after correction for lensing gain. This is surface area 
in the {\it source plane}, or observed surface area corrected for lensing effects. 
It is seen that, as a result of lensing, it becomes possible to detect sources
below the limits achievable in the absence of lensing. The flux-density
domains in which these fainter sources become available are indicated by
the hatched regions in the figure. However, to properly appreciate the
benefit of the lensing survey it is necessary to consider, not simply
the area of sky accessed to a given depth, but to fold in the number
density of sources as a function of flux, in order to highlight the
significant population of fainter sources made accessible by 
lensing. The result of this combination is represented by the shaded
regions of Figs.\ref{how_deep1}.e and \ref{how_deep1}.f, for the example of Abell 2218 and
for the full three-cluster survey, respectively.  To arrive at (e) and (f) the area accessed 
to a given flux has been folded with the 15\,$\mu$m source density as a
function of flux (i.e. with dN/dS).
The subset of the overall source population made accessible by lensing is very 
significant in comparison with the subset of the population falling above the 
unlensed sensitivity threshold.

Lensing effects must be
corrected during analysis (see Sec.\ref{sec:lens_correct}) in order to compare 
lensing results with unlensed sky counts (e.g.\ in the Hubble Deep Field and 
the Lockman Hole -- Rowan-Robinson et al.\,1997; Taniguchi et al.\,1997; 
Aussel et al.\,1999a; Elbaz et al.\,1999 \& 2002).

\begin{table*}[!ht]
 \caption{\em Observational parameters and lensing clusters of galaxies
 used in the survey. On-chip integration time (t$_{int}$) was always
 5.04 seconds and the 3\,$^{\prime\prime}$ per-pixel field-of-view was
 used. Sensitivity in $\mu$Jy is quoted at the 80\% (and 50\%) completeness 
 limits, before taking account of the deepening achieved through the
 lensing amplification. In the table, M and N are the number of steps along each
 dimension of the raster. The increment for each raster step is denoted
 by dm and dn. Each raster was repeated k times.
}
  \label{tab:table1}
   \begin{center}
    \leavevmode
     \footnotesize
      \begin{tabular}{ccccccccccc}
       \hline\hline
\noalign{\smallskip}
   Field    & Filter\,$^{\dagger}$ & Reads    &     n Steps     &   dm              &   dn              &   area      &  Max. depth\,($\mu$Jy)          & Done k &  Tot. t  \\
            & $\lambda_{ref}$\,$\mu$m & per step & M \ \ \ N    & $^{\prime\prime}$ & $^{\prime\prime}$ & $^{\prime\prime^{-2}}$ &80\% (50\%) complete  & times  &  sec.    \\\hline
\noalign{\smallskip}
 A2390 &    6.7                    &    13    &   10 \ \ \ 10   &    7              &    7              &      7      &           52  (38)              &   4    & 28488    \\
\noalign{\smallskip}
       &    14.3                   &    13    &   10 \ \ \ 10   &    7              &    7              &      7      &           92  (50)              &   4    & 28493    \\
\noalign{\smallskip}
\noalign{\smallskip}
 A2218 &    6.7                    &    14    &   12 \ \ \ 12   &   16              &   16              &     20.5    &           79  (54)              &   2    & 22300    \\
 \noalign{\smallskip}
       &    14.3                   &    14    &   12 \ \ \ 12   &   16              &   16              &     20.5    &          167 (121)              &   2    & 22300    \\
\noalign{\smallskip}
\noalign{\smallskip}
 A370  &    6.7                    &    10    &   14 \ \ \ 14   &   22              &   22              &     40.5    &           80  (52)              &   2    & 22366    \\
\noalign{\smallskip}
       &    14.3                   &    10    &   14 \ \ \ 14   &   22              &   22              &     40.5    &          293 (210)              &   2    & 22820    \\
        \hline
      \end{tabular}
   \end{center}
{\footnotesize ($\dagger$) 6.7 and 14.3\,$\mu$m are, respectively, the reference
wavelengths of the ISOCAM LW2 (7\,$\mu$m) and LW3 (15\,$\mu$m) filters, which
have widths (5 to 8.5)\,$\mu$m and (12 to 18)\,$\mu$m.}
\end{table*}

 
\section{Observations}
\label{sec:obsred}

\subsection{Raster observations}
\label{sec:obs}

The fields of the well-known gravitationally lensing galaxy clusters
Abell 370, Abell 2218 and Abell 2390 were observed using the
7\,$\mu$m (LW2: 5 to 8.5$\mu$m) and 15\,$\mu$m (LW3: 12 to 18$\mu$m) filters
of ISOCAM (Cesarsky et al.\,1996). These filters employ the CAM long-wavelength
(LW) detector, which was a 32 $\times$ 32 SiGa array. The fields were observed in
raster mode. The parameters for the rasters and the details of the
instrument configurations used are given in Table 1. 

By employing the CAM 3$^{\prime\prime}$ per-pixel field-of-view
(PFOV) and raster step sizes which were multiples of 1/3 of a pixel,
over many raster steps, a final mosaic pixel size of
1\arcsec \ was achieved.  The diameter of the PSF central maximum at the
first Airy minimum is
0.84$\times\lambda$($\mu$m)\arcsec.\, The FWHM is about half
that amount. More precisely, Okumura (1998) 
reports PSF FWHM values of 3.3\arcsec \ at 7\,$\mu$m and 5\arcsec \ at 15\,$\mu$m.
The final 1\arcsec \ pixel size improves resolution and
allows better cross-identification with observations at other
wavelengths. This strategy differs from many other CAM deep surveys
which used the 6\,$^{\prime\prime}$ per-pixel field-of-view [for
example the ELAIS survey (Oliver et al.\,2000; Serjeant et al.\,2000),
the ISOCAM HDF (Rowan-Robinson et al. 1997), and the Lockman Hole survey
(Taniguchi et al. 1997)].

The rasters achieve greatest depth around their centres because the
dwell--time per position on the sky is a maximum in the centre of each
raster. For purposes of data analysis and interpretation the
rasters were divided into sensitivity zones. The total time required to execute each
individual raster map is listed in Table~\ref{tab:table1}. For the most
sensitive, most redundant, rasters (on A2390) these are also the
maximum per-pixel dwell-times for the inner regions of each raster,
because the corresponding points on the sky
``saw'' the detector array for all raster
positions. Execution times close to 30000 seconds represent rasters
occupying a full half of an ISO revolution - this being the longest
possible individual observation time with the spacecraft due to
ground-station visibility constraints. Table 1 gives the limiting
sensitivities achieved in each raster {\bf before} correcting for the
effects of cluster lensing. 

Results obtained from a subset of the data employed in this paper have
previously been reported by Altieri et al.\,(1999).  L\'emonon et al.\,(1998) 
reported results obtained from a shallower raster over A2390. Both
groups used the reduction method called PRETI, described in Starck et
al.\,(1999).   Barvainis et al.\,(1999) reported CAM photometry of sources
in the field of Abell 2218. The independent reduction method employed in
the current work therefore enables a comparison of results obtained
with different reduction strategies. As can be seen in Table~\ref{tab:barf},
the results show excellent consistency within the 15 to 20\% quoted
precision.


\begin{table*}[!ht]
 \caption{\em Comparison of 7 and 15\,$\mu$m photometry from this paper
 with corresponding values from L\'emonon et al.\,(1998) and Barvainis et
 al.\,(1999), for the few overlap cases. Column 1 lists the label
 assigned to each source in this paper. Columns 2 and 3 (L
 and B) give the identification numbers from the L\'emonon and Barvainis
 papers respectively. Columns 4 and 7 list L\'emonon et al.\,(1998) 15 and
 7\,$\mu$m  photometry. Column 5 gives the Barvainis et al.\,(1999) photometry,
 and columns 6 and 8 give the photometry found in the tables later in
 this paper.
}
  \label{tab:barf}
   \begin{center}
    \leavevmode
     \footnotesize
      \begin{tabular}{lcccrccr}
       \hline
\noalign{\smallskip}
   Source        & L  &  B  &   L\'emonon       &  Barvainis &  Metcalfe  &   L\'emonon       &  Metcalfe \\
                 &    &     & 15\,$\mu$m        & 15\,$\mu$m & 15\,$\mu$m & 7\,$\mu$m         & 7\,$\mu$m \\
                 &    &     &  ($\mu$Jy)        &  ($\mu$Jy) & ($\mu$Jy)  & ($\mu$Jy)         & ($\mu$Jy) \\ 
\noalign{\smallskip}
\noalign{\smallskip}
 ISO\_A2390\_18  & 4  &     & 350$^{+50}_{-40}$ &            &  382+/-59  & 110$^{+40}_{-60}$ &  103+/-30 \\ 
\noalign{\smallskip}
 ISO\_A2390\_27  & 3  &     & 400$^{+60}_{-50}$ &            &  321+/-58  &                   &  153+/-21 \\
\noalign{\smallskip}
 ISO\_A2390\_28a & 2  &     & 440$^{+60}_{-60}$ &            &  461+/-58  &                   &           \\
\noalign{\smallskip}
 ISO\_A2390\_37  & 1  &     & 500$^{+80}_{-70}$ &            &  524+/-61  & 300$^{+50}_{-40}$ &  361+/-26 \\
\noalign{\smallskip}
\noalign{\smallskip}
 ISO\_A2218\_27  &    & 395 &                   & 1100+/-200 &  759+/-57  &                   &  451+/-42 \\
\noalign{\smallskip}
 ISO\_A2218\_35  &    & 317 &                   &  670+/-200 &  541+/-54  &                   &  100+/-57 \\
\noalign{\smallskip}
 ISO\_A2218\_38  &    & 289 &                   &  850+/-200 &  919+/-61  &                   &  113+/-53 \\
\noalign{\smallskip}
 ISO\_A2218\_42  &    & 275 &                   &  570+/-200 &  445+/-54  &                   &   78+/-70 \\
\noalign{\smallskip}
      \hline
      \end{tabular}
   \end{center}
\end{table*}

\section{ISOCAM faint-source data reduction and analysis}
\label{sec:analyse}

\subsection{General remarks}
\label{sec:gen_rem}

\vskip -2pt
Methods and applications of ISOCAM data analysis directed at
the extraction of the faintest sources have been described by Elbaz et al.\,(1998),
Starck et al.\,(1999), Aussel et al.\,(1999a), and Lari et al.\,(2001).

Here we describe an independent method for extracting
faint sources from ISOCAM data. The method can be readily
applied by any user within the ISOCAM Interactive Analysis (CIA) system
(Ott et al. 1997 \& 2001; Delaney \& Ott, 2002). Many of the steps developed 
have been incorporated into CIA as an indirect result of this work.

This section describes the steps required to achieve the removal of cosmic-ray
induced glitches with first and second-order corrections, the
correction of long-term signal drifts, compensation for short-term
responsive transients, faint-source extraction and Monte-Carlo
simulations to calibrate the faint-source data and assign completeness
limits to the survey, and the method employed to correct the results
for the effects of the lensing amplification of the cluster.

\begin{figure}[!hb]
\includegraphics[width=\columnwidth,bb=65 359 559 720]{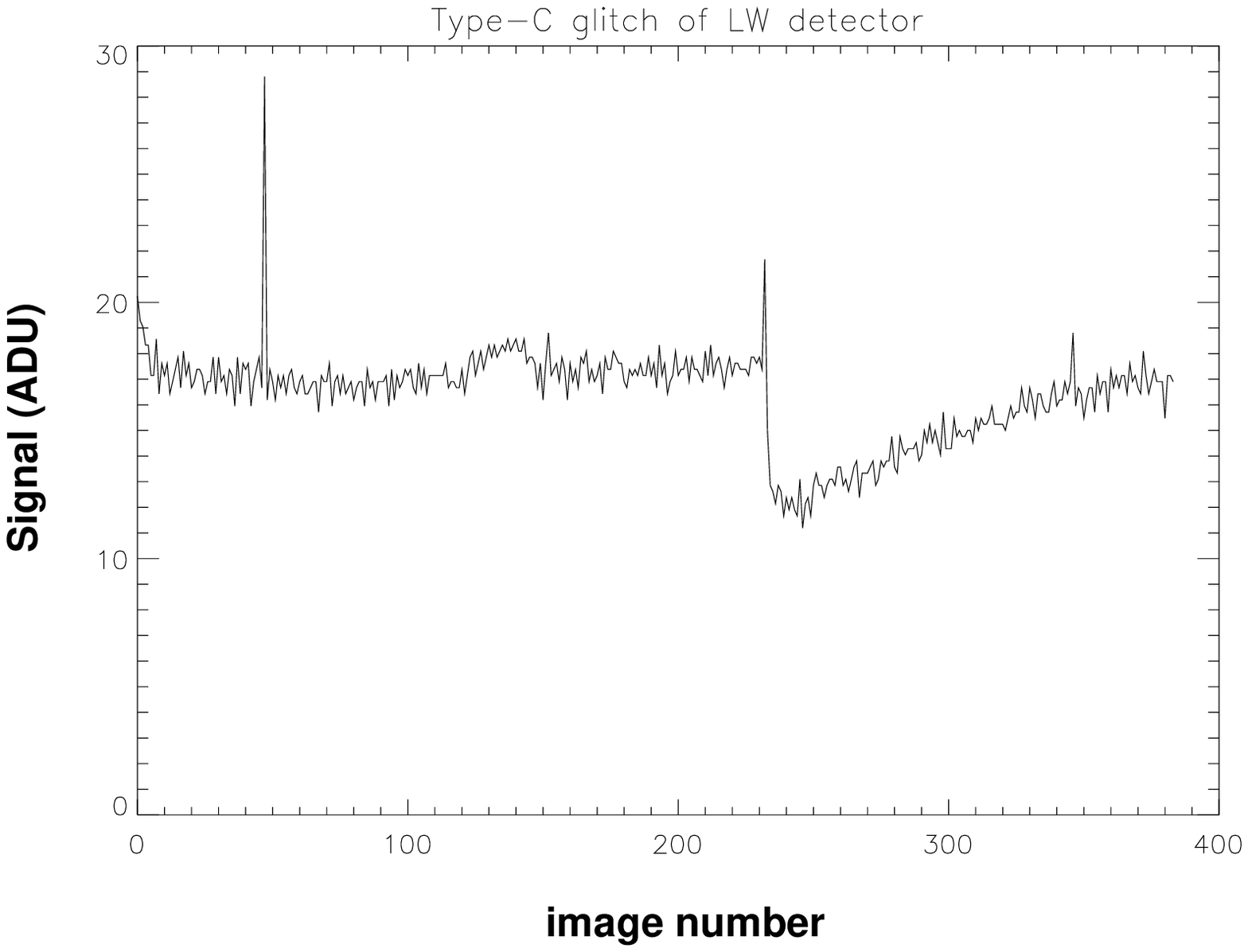}
\caption[]{A typical time history for readout of a CAM LW detector
pixel after primary deglitching. From the start of the sequence at image 0 the signal is seen to
be subject to complex drifts. The spike around image 46 is the residue
of a cosmic-ray-induced glitch left after initial deglitching. Another
glitch residue near image 230 is followed by a severe dip in signal
which recovers slowly (a ``dipper'' glitch). Between images 120 and 150 a
faint source is detected. Its small, upward, signal growth-curve is
characteristic of the short term responsive transient which affects
sources at all flux levels in varying degrees. Truly faint sources,
such as most of the sources discussed in this paper, are not
immediately  discernible in this way above the noise of such a single pixel data
stream.  This figure gives an indication of the challenges facing
the efforts to accurately extract the faintest  sources. Well
established techniques exist to meet these challenges.
}
\label{glitch}
\end{figure}

ISOCAM raw data takes the form of a cube comprised of large numbers of
individual detector image frames, each frame corresponding to one
readout of the full 32x32 detector array. A raw data cube may contain
frames associated with more than one instrument configuration. The
process of decomposing the data into configuration-specific structures
is referred to as "slicing" and is a standard reduction step provided
for in the CAM Interactive Analysis system (Delaney \& Ott, 2002; Ott et al. 1997). 
After slicing, one observation in one
instrument configuration again takes the form of a cube of readouts. In
the case of a raster measurement this cube can be decomposed into
sub-units associated with each raster  position. A source seen, perhaps
repeatedly, by the detector as it rasters over the sky, will then
appear in the data cube as pulsed signal buried in the data stream. 
The signal in an individual detector element, followed throughout the
raster, may then typically take the form of a stream of signal
recording the sky background, combined with dark-current, shot noise,
irregular spikes induced by cosmic ray impacts (glitches), and possibly
containing the quasi-periodic signature of sources seen during the
raster. Even under stable illumination the signal from a given detector
element may drift in time in a manner at least partly correlated with
the illumination history of the array element. This occurs as the response
of the detector element stabilises following any flux step associated
with the start of the observation. Similar response drifts follow
cosmic-ray induced glitches (see Fig.~\ref{glitch}). Apart from the
source-flux-induced drift behaviour, which can be modeled and
corrected (Coulais \& Abergel, 2000; Delaney \& Ott, 2002),  a
lower level drift occurs at a few percent of the background signal. 
This long-term transient must be corrected by purely empirical means.
The  cumulative behaviour is manifested differently for each element of
the detector array,  since the elements differ in intrinsic sensitivity
(i.e. the flat-field) and in their illumination and glitch histories.

It follows that the reduction of ISO faint source data is extremely 
challenging and many work-years have been invested in developing
effective and reliable reduction techniques.  The ultimate performance is
particularly sensitive to the effectiveness with  which the detector's
global responsive transient and the cosmic-ray-induced glitches can be
removed from the data.

\subsection{Initial reduction and first order deglitching}
\label{sec:initial_red}

The data reduction begins as follows: after slicing, to render the
data into instrument configuration-specific structures, the dark
current is subtracted using a time-dependent dark model which takes
account of a well characterised secular variation of current throughout
the orbit (Biviano et al. 2000; Roman \& Ott, 1999). The data-stream from 
each detector
element is then deglitched using an iterative sigma clipping method
which takes advantage of the fact that, for faint source work, the 
uniform celestial background flux dominates the measured signal while
sources are very faint. Abrupt spikes and troughs in the data
stream reveal the occurrence of glitches (rather than sources seen
repeatedly (cyclically) during the raster). 

Glitches are an important, and often dominant, limiting factor for all
aspects of ISOCAM data analysis. However, primary glitch spikes often stand far above the
already high background signal level. Sigma clipping around the average
value of readouts of a pixel with such a perturbed history leads to a
skewed clipping, because the mean value used as reference is severely
perturbed.  



For relatively stabilised (within a few percent of the asymptote) data,
typical of the long-duration data sets used in this paper, the ``Best
Iterative Sigma Clipping Deglitcher'' (BISC) was developed. It applies an
iterative method consisting of four basic steps:

\begin{enumerate}
\item for each raster-position, the flux per pixel is estimated
\item the data vector, which covers many raster positions, is normalised
      by subtracting this flux estimate, reducing the effect of signal
      drifts in comparison with glitches
\item the variance of the normalised data vector is computed
\item outliers are identified by sigma clipping
\item a new (and now improved) estimate of the flux per pixel is made,
      masking the outliers, and the process is repeated until the flux
      level of each pixel in each raster-position reaches a stable
      value.
\end{enumerate}

\noindent Clipping of the original data vector based upon its mean and standard
deviation, with the glitched elements masked, results in a good
rejection of glitch spikes with minimum loss of valid data points.

Very long glitches, referred to as faders and dippers (see 
Fig.~\ref{glitch}), can induce long-lasting signal gradients seen
strongly in the background signal. To
identify and reject these, the mean and variance of the slope per
pixel and raster-position are computed. Data showing a consistent
upward or downward trend are rejected as being of low quality.

\subsection{Removal of short and long--term transients}
\label{sec:short_trans}

After this first-order deglitching the data cube is flat-fielded using
the median frame from the entire raster, but excluding the initial
strong responsive transient which follows the establishment of the
observing configuration.  At this stage an inspection of the data
stream from an individual detector element will typically show a
smoothly varying signal with residual glitch spikes and troughs (see
Fig.~\ref{glitch}). These two remaining characteristics, temporal
variation and residual glitches, represent the fundamental challenge
to the successful extraction of faint point sources.

The temporal drift is decomposed as follows : (a) a ``short-term''
responsive transient following changes in incident illumination on a
detector element (e.g. the onset of a source - see Fig.~\ref{glitch}), 
(b) short term responsive perturbations following cosmic-ray induced 
glitches, \& 
(c) a long-term responsive transient in the measured background,
at a level of a few percent of the background.


The three sources of drift are corrected as follows :

\noindent (a) "Short-term" responsive transients 

The responsive transient which occurs, for example, whenever a source
moves onto a new pixel causes the source signal as recorded by that
pixel to appear roughly like a growing exponential curve (in
appearance, somewhat like the charging-curve of a capacitor, though not
having the same functional form). This response will, in general, not
have stabilised before the raster carries the source on to some other
detector pixel. The result is that the cumulative signal from that
source over the full raster will be suppressed relative to the value it
would have if the signal were allowed to stabilise on each detector
pixel.  This effect can be corrected by applying a physical model to
the data--stream for each detector pixel. However, such application causes a reduction in the
S/N in the final raster  mosaic. For this reason we reduce the rasters 
without applying the model correction for the short-term transient, and
so recover the maximum number of faint sources. We then separately
reduce the data applying the transient model and compare the results
achieved in this way with those obtained without transient modeling,
so deriving a global correction factor to apply to all sources in the
non-transient-corrected map.

\noindent (b) The correction for short term responsive perturbations following 
cosmic-ray induced glitches, and other glitch residues is discussed
in Sec.\ref{sec:kill_glitch}, below.
    
\noindent (c) "Long-term" responsive transients

The low-level long-term signal drift is not significant in terms of
photometric accuracy. However, it leads to an apparent illumination
gradient of one or two percent of the background across the
reconstructed raster maps, and this is sufficient to swamp faint point
sources given the high background characteristic of most ISO
bandpasses.  We have found entirely satisfactory results removing the
long-term drift by simply fitting the signal-stream from each detector
pixel with a broad running median, and then subtracting the smoothed
signal history from the unsmoothed version. The signal-stream from each
detector pixel consists of many readouts of the pixel, with typically
10 to 15 readouts corresponding to one raster position (Table 1).  The
filtering suppresses the long-term drifts and any large scale extended
structure. However, provided the width of the running-filter is several
times the number of readouts  associated with a raster position the
record of point sources in the  data-stream is preserved. We describe
below extensive Monte-Carlo  simulations which allow us to compensate
for any residual perturbations  of photometry which might arise due to
this filtering procedure, and due to the overall processing chain in
general. 

After subtraction of the smoothed time history from the unsmoothed
data the signal history of each detector pixel takes the form of 
noise scattered about zero, interspersed with weak point-source
detections and glitch residues.

\subsection{Removal of glitch residues - Second-order deglitching}
\label{sec:kill_glitch}

First-order deglitching, described above, removes glitch spikes from
the data-stream generated in individual detector pixels. That
deglitching process is not perfect. It may leave residues of glitches
in the data-stream which can resurface in the final raster mosaic and
mimic faint point-sources. The first-order deglitching takes no
account of the fact that a given celestial source will, in general, be
seen by many detector pixels in the course of a full raster
observation.  The number of times a source is seen by different
detector pixels is referred to as the raster redundancy for that
source. The fact that actual sources are multiply detected while
glitches are isolated events allows discrimination of real faint
sources from glitch residues.

The data cube is reorganised as follows to facilitate this
discrimination.  For each raster position the corresponding detector
readouts are averaged to yield a single image. A large data cube is
created having the X and Y dimensions corresponding to the full raster
mosaic on the sky. In the end, each (x,y) coordinate  of this cube
represents one pixel on the sky. The Z dimension is the number of
raster positions in the raster map. The raster-position-images are then
placed into this "sky-cube" in such a way that the first plane of the
sky-cube holds the image from the first raster position, the second
plane of the sky-cube holds the image from the second raster position
displaced laterally in the cube by an amount equal to one raster step,
the third plane of the sky-cube holds the image from the third raster
position, and this process is continued until the full sky-cube is
produced. The corresponding Z track through the cube, for pixel (x,y),
traces each of the occasions on which that sky-pixel  was seen by some
detector pixel. If the raster caused the detector to  move completely
off some patch of sky then the corresponding part of the sky cube will
contain zeros or undefined values which will be ignored in further
data reduction.

If one chooses some (x,y) in the sky map, the
corresponding vector defined by looking down the Z-axis of the cube is
a time history of samples for that sky-pixel. The Z vector will
consistently record enhanced signal when that sky pixel corresponds to
a source. Glitch residues appear as spikes on top of this constant
source signal and can be efficiently rejected by sigma-clipping. Sigma
clipping with a two-sigma threshold is possible without having any
significant effect on the source photometry because all samples for
such a sky-pixel represent flux measurements on the same source. The
extensive Monte-Carlo simulations validate and calibrate this process.

Once all of the above reduction steps have been completed the raster
data cube can be collapsed to produce the final raster map of the
target field.

\subsection{Point-source extraction}
\label{sec:find_source}

Sources listed in this paper fall into two categories:
(i) sources included in the photometric lists, and 
(ii) sources more stringently screened for inclusion in the statistical
source counts. 

\noindent Case (i): for inclusion only in photometric lists detections are
accepted as real which either (a) are confirmed by eye on at least 
two separate IR maps (almost always the case for sources listed in this
paper) or which (b) in the case of a very few sources near the noisy edges
of the map, may fall off the edge of one map due to small deliberate 
mutual displacements of successive rasters on the same field, but which
have clear optical counterparts. All sources listed in this paper have 
optical counterparts. In a few cases these are quite faint - at the limits
of the deepest ground-based observations.
\\

\noindent Case (ii): it was necessary that point-sources to be used for source
counting be extracted by automatic and objective means in order to 
impart well defined statistical properties to the sample. 

The following method was adopted to extract the list of sources for the 
source counts, and to characterise the statistical properties of that
sample:
\vspace*{-0.2cm}

\begin{itemize}
\item First, the real sources in a given field were identified
      independently of the automatic extraction process by identifying
      as real those sources which were visually confirmed on at least
      two independent raster maps, and which have optical
      counterparts.  Not all sources identified in this way are
      counted!
\item The parameters of the automatic source extraction tool,
      SExtractor (Bertin \& Arnouts, 1996) were then tuned for each
      map to reproduce as closely as possible the above set of real
      sources.  Only the set of sources detected in this manner are
      available for source counting.  Sources not found using the 
      chosen settings of SExtractor (only a few percent of the real
      sources) are listed  among the uncountable sources at the end of
      each photometric table later in this paper. The small attrition
      is not a problem, indeed it must be accepted in order to benefit
      from the intrinsic objectivity of the automatic process. It
      simply means  that the source extractor is tuned to detect some
      signal levels of the underlying  source population with less than
      100\% completeness. 
\item Fake source Monte-Carlo simulations (described in detail in the following section)
      are then performed to calibrate the completeness characteristics of the 
      automatically and rigourously selected source sample as a function of source brightness, 
      and in the process to characterise the signal transfer function of the data reduction
      algorithm.  
\end{itemize}  
\vspace*{-0.4cm}

\subsection{Monte-Carlo ``fake-source'' simulations}
\label{sec:monte_carlo}

In what follows we will use the terms ``real'' data to designate data 
recorded during the actual in-flight measurements.  ``Fake'' sources 
are model point sources inserted into the real data for purposes
of simulation. A detection derived from an actual source on the sky is termed 
a real source, as defined in the previous section.  An apparent detection
derived from a cosmic-ray glitch or noise spike is termed a ``false'' detection.
False detections are rejected with high efficiencey because (a) they do not
meet the criteria specified above for identification as real sources, and
(b) they do not correspond to any known, inserted, fake source.

\noindent The Monte-Carlo simulations address two questions:
\vspace*{-0.4cm}

\begin{itemize}
\item What fraction of a well defined source population is recovered 
      by the source extractor to given signal limits. i.e. what is the
      completeness of the survey to different signal limits. 
\item how does signal in the final map relate to signal in the raw data.
\end{itemize}  
\vspace*{-0.4cm}

\noindent Extensive fake-source simulations were performed as follows to address
the above issues:

Artificial point sources, which were filter-specific models of the CAM PSF (Okumura 1998), 
were inserted into the individual detector readouts of the raw data cubes
for the observations. These fake sources were inserted into regions of
the cube free of real sources.  The positions of the fake sources within the
source-free regions of the data frames for a given simulation were determined
by a random number generator. For any given raster observation a total of 
typically one thousand fake sources were inserted into the raw data through 
100 independent simulations.

For each individual simulation, ten fake sources were inserted into the
raw raster cube, all with the same signal level. Then the data
were reduced through the full reduction process, including automatic
source extraction.  Fake sources and real sources face exactly the same
reduction steps.

The simulation was then repeated for ten new fake sources at new positions. This
process was repeated for typically ten ensembles of sources and further
repeated for typically ten to twenty different signal levels of the sources.

For each fake-source signal level in the raw data the
fraction of signal recovered during source extraction on the final map was determined.
This allows recovered signal to be correlated with signal in the raw
data cube over the full range of signal levels spanned by the real sources.

For each fake-source signal level in the raw data the
fraction of the total number of inserted fake-sources that were
recovered during the source extraction process was determined, so establishing
the completeness of the map to that signal level.

\subsection{Correlation of detector signal with actual flux density}
\label{sec:sky_signal}

Now in each of the above cases the properties of sources in the final 
map are correlated, through the simulation, with source signal in the 
raw data, rather than with source brightness on the sky. To establish
the correlation with source flux on the sky, as opposed to signal in
the detector, the transformation from source signal in the raw
data, measured in ADU (Analogue-to-Digital Units), to mJy on the sky, is needed.

We treat this transformation as having two components as follows. 
First, the ISOCAM instrument teams determined through the mission the 
filter-specific transformation from ADU to mJy for a source which has been
seen by the detector long enough for its signal to have stabilised.
The CAM Interactive Analysis (CIA) system (Delaney \& Ott, 2002) contains
these factors, and the ISO Handbook Volume III (CAM) (Blommaert et al.\,2001)
describes and lists them for all CAM filters, as derived from hundreds of calibration
measurements throughout the ISO operational mission.

Second, since our detections refer to sources which have been seen by any given
detector pixel for on-the-order-of only ten readouts, their signals have not stabilised 
completely. The inserted fake-sources do not attempt to model this transient 
response behaviour, therefore a scaling factor must be applied, on top of the
standard calibration factor just described, in order to correctly relate source
signal in the raw data to mJy. The method for determining this additional scaling 
factor has been described in Sec.\ref{sec:short_trans} above 
(Coulais \& Abergel, 2000).

In the end, if n ADU per unit electronic gain per second of point-source
signal is detected within the photometric aperture on the final map, this is
related to source strength in mJy by the transformation : $S = n \times f \times R \times T $

\begin{itemize}
\item n = source signal in detector units (ADU/gain/second) derived by 
          aperture photometry in the final map (a 9 arcsecond aperture was 
          used in all cases)
\item f = the photometric scaling function, determined by simulations.
          It relates measured signal in a photometric aperture to total
          point source signal
\item R = the standard photometric scaling factor relating signal in the
          detector for stabilised data to flux on the sky and taken from 
          the ISOCAM Handbook as described above [units mJy/(ADU/g/s)]
\item T = responsive transient scaling factor: i.e. the ratio [stabilised\_signal/unstab\_signal]
          for the particular measurement in question, derived by reference to
          a physical model
          (see Sec.\ref{sec:short_trans} (Coulais \& Abergel,
	  2000)).
\end{itemize}

\subsection{Correction for the gravitational lensing amplification}
\label{sec:lens_correct}

As has been discussed earlier in this paper, the advantage of using
clusters  of galaxies as natural telescopes is that it becomes possible
to probe fainter  and more distant sources than would otherwise be
reachable in the same  time with standard techniques.  However, source
fluxes, and the effective area of sky surveyed, have to be corrected
for the effects of the lensing amplification.  The necessary
corrections require a good knowledge of the mass distribution of the
clusters based on multiple images of background sources with measured
redshift. These allow accurate probing of the mass distribution of the
lensing cluster. Detailed lensing models of the three
clusters used here have been computed (Kneib et al.\,1996, Pell\'o
et al.\,1999, and B\'ezecourt et al.\,1999).  The correction technique
is now well tested and has been applied on different catalogues of
faint sources, {\it e.g.} by Blain et al.\,(1999) and by Cowie, Barger \& Kneib
(2002) on the submillimetre faint lensed galaxy population, and by Smith et al.\,(2002) 
on the lensed ERO galaxy population.

We face a further complication in determining the area surveyed to a
given flux. In addition to the fact that the
lensing-amplification-corrected detection sensitivity depends on 
position within the map, the {\bf apparent} detection sensitivity  is
also non--uniform and depends upon position in the final image.
This  is a consequence of the rastering observing strategy needed to
cover  sufficient area while obtaining a good flat-field. This effect
is not  critical for wider field observations, like the Abell 370
case, but  becomes important for the Abell 2390 observation, where
the number of  measurement samples per sky position varies strongly
from the centre to the edge of the map. However, this variation in
measurement sensitivity is highly regular over  the map due to the
uniform geometrical character of the raster, and  the Monte Carlo
simulations, which probe sensitivity from region to region within the
maps, suffice to fully take it into account. 

In addition to the above, the redshift distribution of the observed
mid-Infrared  sources is broad, with available measured redshifts ranging up to z=2.8 
and with the redshift
distribution of the lensed sources peaking around  z $\simeq 0.7$. The value of
the  lensing amplification depends on redshift. For those sources for
which  a measured redshift was not available, we have employed the
method described in Sec.\ref{sec:redshift} to sort the sources into
cluster and non-cluster objects, and for the sources deemed to be
background sources the median redshift of the
sample, $z\,\sim\,0.7$, has been assigned.  
\footnote {Altieri et al.\,(1999) showed
that such an approximation, though affecting  slightly the lensing
amplification estimate for a source, does not significantly  impact
the resulting LogN-LogS plot. Concerning the value of 0.7 chosen: in fact, 
the median redshift for known-redshift lensed 
sources (i.e. background sources) in this sample is 0.65. The median
value for all non-cluster field sources is 0.6. But Elbaz et
al.\,(2002) report a value of  $\sim$0.8 for ISOCAM deep survey
sources in general, and so we have chosen a globally  representative
default value still well matched to our sample.  See
Sec.\ref{sec:discuss_counts_plots} for a discussion of  the lower value of
median redshift found in this work.} 

 
To compute the source counts corrected for the effects of lensing,  the
lensing amplification is computed for each source using the cluster 
mass model in order to determine the unlensed flux density of the
source.  Then we compute in the corresponding source plane the area
surveyed at the completeness limit (flux limit) applicable to the source
measurement, taking into account the fact that the detection
sensitivity on the sky (in the image plane) is not uniform due to the
rastering measurement. I.e. different sensitivities apply to different
regions of the maps, so some regions are not viable at some flux
levels.  The inverse of the area surveyed for a given source (a given flux level), 
multiplied by the completeness correction for that flux level, gives the source 
count for that lensing corrected source flux.  Binning and summing these 
results allows us to derive the differential and cumulative source counts 
corrected for the effects of lensing. The cumulative  unlensed source counts 
are presented and discussed in Sec.\ref{sec:counts}.

\section{Results - Images and Photometry}
\label{sec:results_ims}

\vskip -2pt


In this section we present maps of the fields observed showing IR contours at 7
and 15\,$\mu$m overlaid on optical images, and also tables of the 7
and  15\,$\mu$m sources detected for each of the three cluster
fields, giving measured coordinates and flux densities. The coordinates
are those measured by ISOCAM, and may have an apparent, cluster/filter
dependent, systematic offset of a few arcseconds from the optically
determined coordinates of their counterparts. 

\subsection{Cluster images at 7 and 15\,$\mu$m}
\label{sec:images}

The areas of sky covered in the cluster fields are listed in Table 1.
The mid-infrared sources are labeled in each map in order of increasing
RA. The noise level increases towards the edges of the ISOCAM maps.

Figures~\ref{a370lw2}, \ref{a2218lw2} and \ref{a2390lw2} show the
7\,$\mu$m deep images of the three lensing cluster fields.
Figures~\ref{a370lw3}, ~\ref{a2218lw3} and ~\ref{a2390lw3} show the 
15\,$\mu$m maps. Spatial resolution of the ISO maps is sufficiently
good to allow identification of IR correlations with the optical
morphology of objects. For example the ISO source ISO\_A2390\_28a in
Fig.~\ref{a2390lw3} matches the bright optical knot in the straight arc
(Pell\'o et al.\,1991)  near the centre of the A2390 map. From these
images we can remark that  the optical  arc(lets) are generally not
detected at mid-infrared wavelengths. Yet  large numbers of mid-infrared sources that
spectroscopic, photometric or lensing redshifts show to be behind the
lens are detected. Thanks to our comparatively high-resolution images
the correlation of ISO sources with (sometimes extremely faint)
optical counterparts in deep NIR and optical images (e.g. from
HST/WFPC2 or ground-based telescopes) has been quite straightforward. 
We found that at 7\,$\mu$m the bulk of the sources are cluster
galaxies, the rest are lensed background sources, field sources, or
stars. Most 15\,$\mu$m  sources for which spectroscopic
redshifts are available are found to be lensed background
galaxies. The photometric and lensing-inversion
redshifts for the remaining sources, where available, are consistent
with this conclusion.  At 15\,$\mu$m, the cluster-core essentially disappears.
\footnote {But note that this is not true of all galaxy clusters.  
Fadda et al.\,(2000), and Duc et al.\,(2002) found large numbers of 
7\,$\mu$m and 15\,$\mu$m sources in the z=0.181 cluster A1689. 
Coia et al.\,(2003) found a somewhat similar result for the cluster 
Cl\,0024+1654.  (Coia et al.\,2003 used the data analysis methodology 
reported in this paper.)}


\begin{figure*}
\centering
\epsfig{file=3463_f4_c.ps,width=\textwidth,clip=,bb=17 17 593 600}
\caption[]{Contours of the ISOCAM 7\,$\mu$m image of Abell 370 (z = 0.37)
overlaid on a deep CFHT I-band image (also used in figure 5). Sources are
labelled in order of increasing RA. The + sign identifies sources
detected at 7\,$\mu$m but not at 15\,$\mu$m.  If the ID number of a source is enclosed
within a square then the source has been detected at X-ray wavelengths in Chandra data (Bautz et al.\, 2000).
If the ID number of a source is enclosed within a circle then the source has been detected at submm wavelengths
with SCUBA (Ivison et al.\,1998; Smail et al.\,2002; Cowie et al.\,2002). 
The giant arc (near foreground source 11b) is barely detected at 7\,$\mu$m, but the two
giant ellipticals (11c and 12) at the centre of A370 are both
detected.  Source 18 falls off this map and its
approximate location is simply indicated by an arrow. Axes are labelled
in arcseconds.  North is to the top and East to the left. The ISO raster was centred at 
approximately R.A. 02 39 52.9  DEC -01 35 05 (J2000).}
\label{a370lw2}
\end{figure*}
\clearpage

\begin{figure*}
\centering
\epsfig{file=3463_f5_c.ps,width=\textwidth,clip=,bb=17 17 593 600}
\caption[]{Contours of an ISOCAM 15\,$\mu$m image of Abell 370 (z = 0.37)
overlaid on a deep CFHT I-band image. The sources are labeled in order
of increasing RA. The - sign identifies sources detected at
15\,$\mu$m but not at 7\,$\mu$m.  If the ID number of a source is enclosed
within a square then the source has been detected at X-ray wavelengths in Chandra data (Bautz et al.\, 2000).
If the ID number of a source is enclosed within a circle then the source has been detected at submm wavelengths
with SCUBA (Ivison et al.\,1998; Smail et al.\,2002; Cowie et al.\,2002). The giant arc (source number 11)
is well detected by ISO. Some of the mid-IR sources have quite faint
optical counterparts.   Axes are labelled in arcseconds. North is to
the top and East to the left. The ISO raster was centred at approximately 
R.A. 02 39 52.9  DEC -01 35 05 (J2000).}
\label{a370lw3}
\end{figure*}
\clearpage


\begin{figure*}
\centering
\epsfig{file=3463_f6_c.ps,width=\textwidth,clip=,bb=17 17 593 600}
\caption[]{Contours of an ISOCAM 7\,$\mu$m image of Abell 2218 (z = 0.18)
overlaid on a deep Palomar 5m I-band image (Smail, private
communication). The sources are labeled in order of increasing RA. The
+ sign identifies sources detected at 7\,$\mu$m but not at
15\,$\mu$m.  If the ID number of a source is enclosed
within a square then the source has been detected at X-ray wavelengths in Chandra archival images.
Axes are labelled in arcseconds. North is to the top and
East to the left. The ISO raster was centred at approximately 
R.A. 16 35 55.7  DEC 66 12 30 (J2000).}
\label{a2218lw2}
\end{figure*}
\clearpage
 
\begin{figure*}
\centering
\epsfig{file=3463_f7_c.ps,width=\textwidth,clip=,bb=17 17 593 600}
\caption[]{Contours of an ISOCAM  15\,$\mu$m image of Abell 2218 (z = 0.18)
overlaid on a deep Palomar 5m I-band image (Smail,  private
communication). The sources are labeled in order of increasing RA. The
- sign identifies sources detected at 15\,$\mu$m but not at 7\,$\mu$m.
If the ID number of a source is enclosed within a square then the
source has been detected at X-ray wavelengths in Chandra archival
images. Note the strong  ISO detection of the giant arc (number 38),
the several mid-infrared sources  corresponding to faint optical
counterparts [e.g. 43, 67] and the fact that the IR  ``arclets'' are
generally not from the same  population as the known optical arclets.
Source 27 is a foreground galaxy. Axes are labelled in arcseconds.  
North is to the top and East to the left. The ISO raster was centred at
approximately  R.A. 16 35 55.7  DEC 66 12 30 (J2000).}
\label{a2218lw3}
\end{figure*}
\clearpage



\begin{figure*}
\centering
\epsfig{file=3463_f8_c.ps,width=\textwidth,clip=,bb=17 17 593 600}
\caption[]{Contours of an ISOCAM 7\,$\mu$m image of Abell 2390 (z = 0.23)
overlaid on a KPNO R-band image (Soucail, private communication). The
sources are labeled in order of increasing RA. The + sign identifies
sources detected at 7\,$\mu$m but not at 15\,$\mu$m. If the ID number
of a source is enclosed within a square then the source has been
detected at X-ray wavelengths in Chandra data (Fabian et al.\,2000;
Wilman et al.\,2000a). If the ID number of a source is enclosed within a
circle then the source has been detected at submm wavelengths with
SCUBA (Smail et al.\,2002; Cowie et al.\,2002). Axes are
labelled in arcseconds. North is to the top and East to the left.
The ISO raster was centred at approximately R.A. 21 53 33.9  DEC 17 42 00 (J2000).}
\label{a2390lw2}
\end{figure*}
\clearpage

\begin{figure*}
\centering
\epsfig{file=3463_f9_c.ps,width=\textwidth,clip=,bb=17 17 593 600}
\caption[]{Contours of an ISOCAM 15\,$\mu$m image of Abell 2390 (z = 0.23)
overlaid on a KPNO R-band image (Soucail, private communication). The
sources are labeled in order of increasing RA. The - sign identifies
sources detected at 15\,$\mu$m but not at 7\,$\mu$m. If the ID number
of a source is enclosed within a square then the source has been
detected at X-ray wavelengths in Chandra data (Fabian et al.\,2000;
Wilman et al.\,2000a). If the ID number of a source is enclosed within a
circle then the source has been detected at submm wavelengths with
SCUBA (Smail et al.\,2002; Cowie et al.\,2002). Note the ISO  detection
of part of the straight arc near the centre of the image (source number
28a), and the two strong ISO detections of the optically very faint
sources 18 and 25.  These objects have extremely red colours and very
faint visual fluxes.  Axes are labelled in arcseconds. North is to the
top and  East to the left. The ISO raster was centred at approximately
R.A. 21 53 33.9  DEC 17 42 00 (J2000).}
\label{a2390lw3}
\end{figure*}
\clearpage


Some ISO targets have extremely red optical and NIR colours.  It
appears that  15\,$\mu$m imaging favours selection of star-forming
galaxies and dusty AGNs that are not evident in UV/optical surveys.

\vskip -2pt
\subsection{Photometry}
\label{sec:tables}

Table~\ref{tab:limits} lists, for all clusters and both bandpasses,
the sensitivities achieved to three completeness levels (90\%, 80\% and 50\%) 
in each of the sensitivity zones defined for each raster. The areas of
the zones are given in square arcseconds. The zones are labeled  a, b, c, 
and d, and represent steps of a factor of 2 in observation dwell--time 
per sky pixel (or raster redundancy per pixel). The low redundancy noisy
``a''-regions were excluded from contributing to the source counts, but contributed
sources to the photometric lists.

\begin{table}[!ht]
 \caption{\em Sensitivities achieved in the survey. For each cluster in each 
 filter the raster is divided into
 zones a to d. Zone d is the inner, most redundant and sensitive region.
 Zone a is the outer, least redundant and sensitive region. The limiting 
 flux for detections with 90\%, 80\% and 50\% completeness, and the area
 covered, is given in square arcseconds for each zone. 
}
  \label{tab:limits}
   \begin{center}
    \leavevmode
     \scriptsize 
      \begin{tabular}{c@{}c@{}ccccc@{}c@{}}
       \hline\hline
        \noalign{\smallskip}
 ~Field~                                        & ~Filter\,$^{\dagger}$~ & Zone & \multicolumn{3}{c}{Depth (mJy)} & Zone~                & Tot. Usable \\
                                              &                      &      &  90\% &    80\% &   50\%            & Area                 &  Area         \\ 
                                              &                      &      &       &         &                   & [$\prime\prime^{2}$] & [$\prime\prime^{2}$]\\ 
\noalign{\smallskip}
\hline
\noalign{\smallskip}
                                              & LW2                  &  a   & 0.163&   0.149&   0.107   &  6135 &            \\
                                              &                      &  b   & 0.087&   0.08 &   0.06    &  5154 &            \\
                                              &                      &  c   & 0.063&   0.059&   0.046   &  6835 &            \\
                                              &                      &  d   & 0.057&   0.052&   0.038   &  7157 &   25281    \\
\begin{rotate}{90}{\hspace*{-0.7cm}\bf Abell 2390}\end{rotate}& LW3  &  a   & 0.475&   0.392&   0.220   &  6125 &            \\
                                              &                      &  b   & 0.415&   0.281&   0.116   &  5192 &            \\
                                              &                      &  c   & 0.140&   0.128&   0.092   &  6804 &            \\
                                              &                      &  d   & 0.107&   0.092&   0.050   &  7160 &   25281    \\ 
\noalign{\smallskip}
\hline
\noalign{\smallskip}

                                              & LW2                  &  a   & 0.294&   0.268&   0.180   &  9728 &            \\
                                              &                      &  b   & 0.180&   0.166&   0.124   & 13564 &            \\
                                              &                      &  c   & 0.112&   0.103&   0.078   & 19004 &            \\
                                              &                      &  d   & 0.087&   0.79 &   0.054   & 31688 &   73984    \\
\begin{rotate}{90}{\hspace*{-0.7cm}\bf Abell 2218}\end{rotate}& LW3  &  a   & 0.449&   0.412&   0.300   &  9608 &            \\
                                              &                      &  b   & 0.328&   0.309&   0.251   & 13319 &            \\
                                              &                      &  c   & 0.266&   0.249&   0.198   & 19100 &            \\
                                              &                      &  d   & 0.186&   0.167&   0.121   & 31957 &   73984    \\
\noalign{\smallskip}
\hline
\noalign{\smallskip}
                                              & LW2                  &  a   & \multicolumn{3}{c}{Insufficient redundancy} & 10187 &            \\
                                              &                      &  b   & \multicolumn{3}{c}{Insufficient redundancy} & 21344 &            \\
                                              &                      &  c   & 0.135&   0.124&   0.09    & 32579 &            \\
                                              &                      &  d   & 0.1  &   0.08 &   0.052   & 81814 &  114393    \\
\begin{rotate}{90}{\hspace*{-0.7cm}\bf Abell 370}\end{rotate}& LW3   &  a   & \multicolumn{3}{c}{Insufficient redundancy} &  9434 &            \\
                                              &                      &  b   & \multicolumn{3}{c}{Insufficient redundancy} & 22358 &            \\
                                              &                      &  c   & \multicolumn{3}{c}{Insufficient redundancy} & 31808 &            \\
                                              &                      &  d   & 0.331&   0.293&   0.21    & 82324 &   82324    \\ 
\noalign{\smallskip}
\hline
      \end{tabular}
   \end{center}
($\dagger$) LW2 and LW3 filters have reference wavelengths 6.7 and 15\,$\mu$m, respectively. 
\end{table}

%

The photometry for the 145 mid-infrared background, cluster and foreground
sources detected in the survey is listed in Tables~\ref{tab:a370lw3tab}
to \ref{tab:a2390lw2tab}. For each table, the significance of the columns 
is as follows :

{\bf Column 1} gives the source identifier. A minus sign beside the
source number indicates a source seen only at 15\,$\mu$m. A plus sign
indicates a source seen only at 7\,$\mu$m. {\bf Column 2} gives the source signal
in detector units with the precision in detector units listed in {\bf Column 3}. 
The square brackets in {\bf Column 4} enclose the
value of the significance of the detection with reference to the local
background noise in the map.  The significance is not the
same as the signal divided by the precision.  The precision includes
factors roughly dependent on source signal (for example due to
cosmic-ray induced glitches which change the detector responsivity) and
which affect the accuracy of the flux determination.  The significance
is the ratio of the source signal to the 1-sigma local noise floor of
the map as derived from the Monte-Carlo simulations (the scatter on the
recovered signal of the faintest fake sources recovered with greater 
than 50\% completeness). The distinction is clearly understood if one considers a
bright source affected by a glitch (limited precision but high
significance). {\bf Column 5} gives the flux density of the source in mJy, 
with the associated precision listed in {\bf Column 6}. {\bf Column 7} gives 
the intrinsic flux density of the source after correction for lensing 
amplification, with the
associated precision listed in {\bf Column 8}, which is the quadrature sum of
the apparent flux precision with an estimate of the accuracy of the
lensing amplification model at the source location. (The absoulte uncertainty
on the intrinsic flux-density is smaller than the absoulte uncertainty on the
apparent flux-density because of the lensing factor, but the fractional error
on the intrinsic flux-density is, of course, increased by the quadrature summation of 
errors.) {\bf Columns 9} and
{\bf 10} list the  Right Ascenscion and Declination in J2000 coordinates, as
determined by the ISO Attitude and Orbit Control System (AOCS). Due to jitter in 
the positioning
of ISOCAM's lens wheel the inherent accuracy of the AOCS
($\sim$3\arcsec\ )
was not always achieved, and these coordinates may differ {\bf systematically} by a
few arcseconds (typically less than 6\arcsec ) from the coordinates of optical
counterparts, with map dependent offsets being needed from map to
map to align the ISO images with the optical data. {\bf Column 11} lists the
spectroscopic redshift for the source, if known, along with a key indicating the 
origin of the z information.  The key is decoded at the end of each
table. Where no redshift information is available
sources are sorted into cluster-member and non-cluster-member 
background-source categories based on their IR colours, as mentioned above in 
Sec.\ref{sec:lens_correct} and described in detail in
Sec.\ref{sec:redshift}.  As already stated in
Sec.\ref{sec:lens_correct}, sources identified as background sources in
this way are assigned a redshift of 0.7, roughly the median value for
the 15\,$\mu$m background population. 

Each 15\,$\mu$m table is organised into three sections vertically, 
separated by blank rows across the table.  The upper section
lists the sources meeting all of the criteria for countability,
{\bf and} detected above the 80\% completeness threshold. The middle section 
lists the sources meeting all of the criteria for countability,
{\bf and} detected above the 50\% completeness threshold. The bottom
section of each table lists the sources rejected for counting
(as cluster members or as stars) or detected below the 50\% completeness threshold.
In all cases the significance of the detection is given.

The 7\,$\mu$m tables are divided into two sections vertically, 
separated by a blank row across the table.  The upper section
lists the sources meeting all of the criteria for countability,
{\bf and} detected above the 80\% completeness threshold. The bottom
section of each table lists the sources rejected for counting
(as cluster members or as stars) or detected below the 80\% completeness threshold.
In every case the significance of the detection is given.

\vskip -10pt
\begin{table*}[!ht]
 \caption{\em 15\,$\mu$m sources detected in the field of Abell 370. The
significance of the columns is explained in the text
(Sec.\ref{sec:tables}). Horizontal blank lines divide the table
into three parts. The upper section holds non-cluster sources detected
above the 80\% completeness limit.  The middle section holds
non-cluster sources detected above the 50\% completeness limit.  The
bottom section holds cluster sources and any non-cluster sources not
meeting statistical criteria for inclusion in the counts. (See
Sec.\ref{sec:find_source} \& Sec.\ref{sec:monte_carlo}.) When
the redshift  is given in italics it has been assigned to 0 for stars,
or to the median redshift of the background population, $\sim$0.7, for
background objects lacking spectroscopic redshift (see
Sections~\ref{sec:redshift} and \ref{sec:discuss_counts_plots}), 
or the source is in the cluster. The
letters associated with z values identify the origin of the z value,
expanded at the end of the table.  Source IDs accompanied by (-) signs are
sources detected at 15\,$\mu$m, but not at 7\,$\mu$m. The mean redshift
of the cluster is z = 0.374.
}
  \label{tab:a370lw3tab}
   \begin{center}
    \leavevmode
     \scriptsize
      \begin{tabular}{lcllllllccl}
       \hline\hline
        \noalign{\smallskip}
 Object ID  &   S    & \multicolumn{2}{c}{Precision}            &   F   & Precision    & Lens corr. & Precision    &   R.A.      &     DEC    &   z$_{spec}$    \\ 
 ISO\_A370\ & (ADU)  & \multicolumn{2}{c}{  (ADU) \&}           & (mJy) & +/-mJy       &  F\,(mJy)  & +/-mJy       &             &            &    if known     \\
            &        & \multicolumn{2}{c}{Significance}         &       & (1$\sigma $) &            & (1$\sigma $) &             &            &                 \\
            &        & \multicolumn{2}{c}{[$S/\sigma_{floor}$]} &       &              &            &              &             &            &                 \\\hline
\noalign{\smallskip}
  01 -      & 0.325  & 0.062 & [8.5 ]                           & 0.361 &  0.280       &    0.325   &    0.070     & 2: 39: 45.6 & -1: 33: 13 & {\it 0.700}     \\
  03 -      & 0.560  & 0.064 & [14.7]                           & 0.656 &  0.080       &    0.615   &    0.077     & 2: 39: 48.1 & -1: 33: 26 & 0.44$^K$        \\
  08 -      & 0.395  & 0.059 & [10.3]                           & 0.449 &  0.074       &    0.449   &    0.074     & 2: 39: 51.7 & -1: 35: 16 & 0.306$^M$       \\
  09        & 0.399  & 0.059 & [10.5]                           & 0.454 &  0.074       &    0.198   &    0.040     & 2: 39: 51.7 & -1: 35: 52 & 2.803$^I$       \\
  10 -      & 0.314  & 0.057 & [8.3 ]                           & 0.347 &  0.071       &    0.331   &    0.070     & 2: 39: 51.7 & -1: 36: 09 & 0.42$^K$        \\
  11 -      & 1.08   & 0.285 & [28.4]                           & 1.300 &  0.360       &    0.200   &    0.060     & 2: 39: 53.0 & -1: 34: 59 & 0.724$^{S8}$    \\
  15        & 0.434  & 0.058 & [11.4]                           & 0.498 &  0.073       &    0.498   &    0.073     & 2: 39: 54.4 & -1: 32: 28 & 0.045$^N$       \\
  16 -      & 0.346  & 0.061 & [9.1 ]                           & 0.388 &  0.077       &    0.357   &    0.071     & 2: 39: 55.9 & -1: 37: 01 & {\it 0.700}     \\
  17        & 1.148  & 0.062 & [30.2]                           & 1.395 &  0.078       &    0.788   &    0.080     & 2: 39: 56.4 & -1: 34: 21 & 1.062$^{S9/B}$  \\
  19 -      & 0.279  & 0.057 & [7.3 ]                           & 0.303 &  0.071       &    0.278   &    0.066     & 2: 39: 56.9 & -1: 36: 49 & {\it 0.700}     \\
  21 -      & 0.705  & 0.285 & [18.6]                           & 0.840 &  0.360       &    0.780   &    0.080     & 2: 39: 58.0 & -1: 32: 26 & {\it 0.700}     \\
  23 -      & 0.318  & 0.061 & [8.4 ]                           & 0.352 &  0.077       &    0.320   &    0.071     & 2: 40: 00.1 & -1: 35: 34 & {\it 0.700}     \\
\noalign{\smallskip}	
\noalign{\smallskip}	
  02 -      & 0.239  & 0.047 & [6.3 ]                           & 0.253 &  0.059       &    0.238   &    0.056     & 2: 39: 46.3 & -1: 36: 48 & {\it 0.700}     \\
  20 -      & 0.218  & 0.047 & [5.7 ]                           & 0.227 &  0.059       &    0.196   &    0.051     & 2: 39: 57.5 & -1: 33: 16 & {\it 0.700}     \\
  24 -      & 0.26   & 0.047 & [6.8 ]                           & 0.280 &  0.059       &    0.258   &    0.055     & 2: 40: 00.6 & -1: 33: 44 & {\it 0.700}     \\
  25 -      & 0.262  & 0.047 & [6.9 ]                           & 0.282 &  0.059       &    0.282   &    0.059     & 2: 40: 00.8 & -1: 36: 11 & 0.23$^M$        \\
\noalign{\smallskip}	
\noalign{\smallskip}	
  04 -      & 0.203  & 0.038 & [5.3 ]                           & 0.208 &  0.047       &    0.152   &    0.035     & 2: 39: 48.6 & -1: 35: 14 & {\it 0.700}     \\
  05 -      & 0.203  & 0.038 & [5.3 ]                           & 0.208 &  0.047       &    0.154   &    0.036     & 2: 39: 49.4 & -1: 35: 33 & {\it 0.700}     \\
  14 -      & 0.384  & 0.059 & [10.1]                           & 0.435 &  0.074       &    0.435   &    0.074     & 2: 39: 54.3 & -1: 33: 30 & {\it 0.375$^M$} \\
  22        & 0.55   & 0.064 & [14.5]                           & 0.644 &  0.080       &    0.644   &    0.081     & 2: 39: 59.0 & -1: 33: 55 & {\it 0 star}    \\

\noalign{\smallskip}                                           
\noalign{\smallskip}
\noalign{\smallskip}
\noalign{\smallskip}
\multicolumn{10}{l}{  Redshift references :  S8 = Soucail et al. 1988; M  = Mellier et al. 1988; I  = Ivison et al. 1998. }        \\ 
\multicolumn{10}{l}{ \ \ \ \ \ \ \ \ \ \ \ \ \ \ \ \ \ \ \ \ \ \ \ \ \ \ \ \ S9 = Soucail et al. 1999; B  = Barger et al. 1999; K  = Kneib, unpublished; N  = NED } \\
\noalign{\smallskip}
\noalign{\smallskip}
        \hline
      \end{tabular}
   \end{center}
\end{table*}

\vskip -10pt

\begin{table*}[!ht]
 \caption{\em 15\,$\mu$m sources detected in the field of Abell 2218. 
The significance of the columns is explained in the text
(Sec.\ref{sec:tables}). Horizontal blank lines divide the table
into three parts. The upper section holds non-cluster sources detected
above the 80\% completeness limit.  The middle section holds
non-cluster sources detected above the 50\% completeness limit.  The
bottom section holds cluster sources and any non-cluster sources not
meeting statistical criteria for inclusion in the  counts. (See
Sec.\ref{sec:find_source} \& Sec.\ref{sec:monte_carlo}.) When
the redshift  is given in italics it has been assigned to 0 for stars,
or to the median redshift of the background population, $\sim$0.7, for
background objects lacking spectroscopic redshift (see
Sections~\ref{sec:redshift} and \ref{sec:discuss_counts_plots}), 
or the source is in the cluster. The
letters associated with z values identify the origin of the z value,
expanded at the end of the table.  Source IDs accompanied by (-) signs
are sources detected at 15\,$\mu$m, but not at 7\,$\mu$m. The mean
redshift of the cluster is z = 0.1756.
}
  \label{tab:a2218lw3tab}
   \begin{center}
    \leavevmode
     \scriptsize
      \begin{tabular}{lcllllllccl}
       \hline\hline
        \noalign{\smallskip}

 Object ID  &    S    & \multicolumn{2}{c}{Precision}            &   F      & Precision    & Lens corr. & Precision    &   R.A.       &     DEC    &   z$_{spec}$       \\ 
ISO\_A2218\ &  (ADU)  & \multicolumn{2}{c}{  (ADU) \&}           &   (mJy)  & +/-mJy       &  F\,(mJy)  & +/-mJy       &              &            &    if known        \\
            &         & \multicolumn{2}{c}{ Significance}        &          & (1$\sigma $) &            & (1$\sigma $) &              &            &                    \\
            &         & \multicolumn{2}{c}{[$S/\sigma_{floor}$]} &          &              &            &              &              &            &                    \\
\noalign{\smallskip}\hline
\noalign{\smallskip}
  02        &  0.363  & 0.016 & 	[21.2]                   & 0.472    &    0.028     &   0.406    &    0.030     & 16: 35: 32.2 & 66: 12: 02 & 0.53$^K$           \\
  08 -      &  0.306  & 0.019 & 	[18]                     & 0.393    &    0.026     &   0.249    &    0.034     & 16: 35: 37.2 & 66: 13: 12 & 0.68$^K$           \\
  13 -      &  0.314  & 0.019 & 	[18.5]                   & 0.404    &    0.026     &   0.222    &    0.037     & 16: 35: 40.9 & 66: 13: 30 & {\it 0.700}        \\
  15a -     &  0.279  & 0.041 & 	[16.4]                   & 0.368    &    0.054     &   0.222    &    0.044     & 16: 35: 42.1 & 66: 12: 10 & {\it 0.700}        \\
  17        &  0.134  & 0.046 & 	[7.9]                    & 0.177    &    0.061     &   0.115    &    0.042     & 16: 35: 43.7 & 66: 11: 54 & {\it 0.700}        \\ 
  19 -      &  0.213  & 0.017 & 	[12.5]                   & 0.264    &    0.024     &   0.208    &    0.024     & 16: 35: 44.4 & 66: 14: 23 & 0.72$^K$           \\
  27        &  0.575  & 0.046 & 	[34]                     & 0.759    &    0.057     &   0.759    &    0.057     & 16: 35: 48.8 & 66: 13: 00 & 0.103$^{L2}$       \\
  35        &  0.41   & 0.041 & 	[24]                     & 0.541    &    0.054     &   0.204    &    0.050     & 16: 35: 53.2 & 66: 12: 57 & 0.474$^{Eb}$       \\
  38        &  0.696  & 0.046 & 	[41]                     & 0.919    &    0.061     &   0.222    &    0.064     & 16: 35: 54.9 & 66: 11: 50 & 1.034$^P$          \\
  42        &  0.337  & 0.041 & 	[20]                     & 0.445    &    0.054     &   0.274    &    0.048     & 16: 35: 55.3 & 66: 13: 13 & 0.45$^K$           \\
  43 -      &  0.349  & 0.019 & 	[20.5]                   & 0.452    &    0.028     &   0.367    &    0.031     & 16: 35: 55.4 & 66: 10: 30 & {\it 0.700}        \\
  46 -      &  0.137  & 0.046 & 	[8.1]                    & 0.181    &    0.061     &   0.144    &    0.049     & 16: 35: 56.6 & 66: 14: 03 & {\it 0.700}        \\
  48        &  0.214  & 0.045 & 	[12.6]                   & 0.283    &    0.059     &   0.248    &    0.053     & 16: 35: 57.2 & 66: 14: 37 & {\it 0.700}        \\
  50 -      &  0.353  & 0.051 & 	[10.7]                   & 0.415    &    0.073     &   0.364    &    0.066     & 16: 35: 57.5 & 66: 10: 08 & {\it 0.700}        \\
  51        &  0.188  & 0.045 & 	[11]                     & 0.248    &    0.059     &   0.132    &    0.038     & 16: 35: 58.0 & 66: 11: 11 & {\it 0.700}        \\
  52        &  0.358  & 0.041 & 	[22]                     & 0.473    &    0.054     &   0.178    &    0.044     & 16: 35: 58.7 & 66: 11: 47 & 0.55$^K$           \\
  53 -      &  0.341  & 0.041 & 	[20]                     & 0.450    &    0.054     &   0.113    &    0.034     & 16: 35: 58.8 & 66: 11: 59 & 0.693$^{Eb}$       \\
  61a       &  0.14   & 0.046 & 	[8.2]                    & 0.185    &    0.061     &   0.136    &    0.061     & 16: 36: 03.6 & 66: 13: 32 & {\it 0.700}        \\ 
  65 -      &  0.657  & 0.031 & 	[38.1]                   & 0.878    &    0.043     &   0.734    &    0.052     & 16: 36: 05.1 & 66: 10: 43 & 0.64$^K$           \\
  67        &  0.27   & 0.041 & 	[15.9]                   & 0.356    &    0.054     &   0.259    &    0.045     & 16: 36: 06.4 & 66: 12: 30 & {\it 0.700}        \\
  68        &  0.221  & 0.044 & 	[13]                     & 0.292    &    0.058     &   0.247    &    0.050     & 16: 36: 07.2 & 66: 12: 59 & 0.42$^K$           \\
  69 -      &  0.137  & 0.046 & 	[8]                      & 0.181    &    0.061     &   0.135    &    0.047     & 16: 36: 07.6 & 66: 12: 21 & {\it 0.700}        \\
  73b -     &  0.284  & 0.046 & 	[8.6]                    & 0.326    &    0.060     &   0.292    &    0.055     & 16: 36: 10.3 & 66: 13: 48 & {\it 0.700}        \\
\noalign{\smallskip}	      
\noalign{\smallskip}	      
  09 -      &  0.168  & 0.017 & 	[10]                     & 0.202    &    0.024     &   0.111    &    0.021     & 16: 35: 39.0 & 66: 13: 07 & 0.68$^K$           \\
  10 -      &  0.098  & 0.037 & 	[5.8]                    & 0.129    &    0.049     &   0.091    &    0.036     & 16: 35: 39.4 & 66: 12: 05 & {\it 0.700}        \\
  26 -      &  0.093  & 0.037 & 	[5.5]                    & 0.123    &    0.049     &   0.056    &    0.025     & 16: 35: 48.2 & 66: 12: 01 & {\it 0.700}        \\
  29 -      &  0.112  & 0.044 & 	[6.6]                    & 0.148    &    0.049     &   0.079    &    0.029     & 16: 35: 49.6 & 66: 13: 34 & 0.521$^{Eb}$       \\
  37 -      &  0.117  & 0.037 & 	[6.9]                    & 0.154    &    0.049     &   0.109    &    0.036     & 16: 35: 54.8 & 66: 10: 53 & {\it 0.700}        \\
  44 -      &  0.101  & 0.037 & 	[6]                      & 0.133    &    0.049     &   0.086    &    0.033     & 16: 35: 55.7 & 66: 13: 28 & 0.596$^{Eb}$       \\
\noalign{\smallskip}	      
\noalign{\smallskip}	      
%
  05 -      &  0.208  & 0.046 & 	[6.3]                    & 0.227    &    0.060     &   0.164    &    0.046     & 16: 35: 34.3 & 66: 13: 11 & {\it 0.700}        \\
  06        &  0.128  & 0.017 & 	[7.5]                    & 0.147    &    0.024     &   0.147    &    0.024     & 16: 35: 34.7 & 66: 12: 32 & {\it 0.180$^{L2}$} \\
  07 -      &  0.203  & 0.046 & 	[6.1]                    & 0.220    &    0.060     &   0.151    &    0.043     & 16: 35: 36.4 & 66: 13: 31 & {\it 0.700}        \\
  20 -      &  0.068  & 0.037 & 	[4]                      & 0.090    &    0.049     &   0.018    &    0.011     & 16: 35: 44.7 & 66: 12: 45 & 0.475$^{Eb}$         \\
  25 -      &  0.103  & 0.037 & 	[6]                      & 0.136    &    0.049     &   0.136    &    0.049     & 16: 35: 48.3 & 66: 11: 26 & {\it 0.1741$^{L2}$}\\
  30        &  0.13   & 0.046 & 	[7.6]                    & 0.172    &    0.058     &   0.172    &    0.058     & 16: 35: 49.7 & 66: 12: 22 & {\it 0.178$^{L2}$} \\
  34 -      &  0.072  & 0.037 & 	[4.2]                    & 0.095    &    0.049     &   0.095    &    0.049     & 16: 35: 53.1 & 66: 12: 30 & {\it 0.179$^{Eb}$} \\
  36a -     &  0.079  & 0.037 & 	[4.6]                    & 0.104    &    0.049     &   0.080    &    0.038     & 16: 35: 54.2 & 66: 14: 05 & {\it 0.700}        \\
  49        &  0.316  & 0.019 & 	[18.6]                   & 0.407    &    0.026     &   0.407    &    0.026     & 16: 35: 57.8 & 66: 14: 22 & {\it 0.15$^{SED}$} \\ 
  55 -      &  0.515  & 0.216 & 	[2.8]                    & 0.686    &    0.282     &   0.637    &    0.262     & 16: 36: 00.3 & 66: 09: 41 & {\it 0.700}        \\
  56        &  0.087  & 0.037 & 	[5.1]                    & 0.115    &    0.049     &   0.115    &    0.049     & 16: 36: 00.7 & 66: 13: 20 & {\it 0 star}       \\
  59 -      &  0.129  & 0.017 & 	[7.6]                    & 0.148    &    0.024     &   0.123    &    0.021     & 16: 36: 02.4 & 66: 13: 50 & {\it 0.700}        \\
  60        &  0.087  & 0.037 & 	[5.1]                    & 0.115    &    0.049     &   0.085    &    0.036     & 16: 36: 04.1 & 66: 13: 03 & 0.2913$^{L2}$      \\ 
  70 -      &  0.144  & 0.017 & 	[8.5]                    & 0.169    &    0.024     &   0.128    &    0.021     & 16: 36: 07.7 & 66: 11: 37 & {0.1703$^{L2}$}    \\ 
  72        &  0.132  & 0.017 & 	[7.8]                    & 0.152    &    0.024     &   0.152    &    0.024     & 16: 36: 09.6 & 66: 11: 57 & {\it 0 star}       \\
  73a -     &  0.190  & 0.046 & 	[5.8]                    & 0.204    &    0.060     &   0.182    &    0.054     & 16: 36: 09.6 & 66: 13: 47 & {\it 0.700}        \\
  74 -      &  0.135  & 0.017 & 	[7.9]                    & 0.156    &    0.024     &   0.134    &    0.021     & 16: 36: 09.9 & 66: 13: 15 & {\it 0.700}        \\
\noalign{\smallskip}
\noalign{\smallskip}
\multicolumn{10}{l}{  Redshift references :  P  = Pell\'o et al. 1991; L2 = LeBorgne et al. 1992; Eb = Ebbels et al. 1998; K  = Kneib, unpublished. } \\
\multicolumn{10}{l}{ \ \ \ \ \ \ \ \ \ \ \ \ \ \ \ \ \ \ \ \ \ \ \ \ \ \ \ \ SED = Spectral fitting. }                             \\
\noalign{\smallskip}
\noalign{\smallskip}
   \hline 
      \end{tabular}
   \end{center}
\end{table*}

\vskip -10pt

\begin{table*}[!ht]
 \caption{\em 15\,$\mu$m sources detected in the field of Abell 2390. 
The significance of the columns is explained in the text
(Sec.\ref{sec:tables}). Horizontal blank lines divide the table into three parts. The upper
section holds non-cluster sources detected above the 80\% completeness limit.  The middle section
holds non-cluster sources detected above the 50\% completeness limit.  The bottom section holds
cluster sources and any non-cluster sources not meeting statistical criteria for inclusion in the 
counts. (See
Sec.\ref{sec:find_source} \& Sec.\ref{sec:monte_carlo}.) When the redshift 
is given in italics it has been assigned to 0 for stars, or to the median redshift of the background
population, $\sim$0.7, for background objects lacking spectroscopic redshift (see
Sections~\ref{sec:redshift} and \ref{sec:discuss_counts_plots}), 
or the source is in the cluster. The letters associated with z values
identify the origin of the z value, expanded at the end of the table. 
Source IDs acompanied by (-) signs are sources detected at 15\,$\mu$m, but
not at 7\,$\mu$m. The mean redshift of the cluster is z = 0.232.
}
  \label{tab:a2390lw3tab}
   \begin{center}
    \leavevmode
     \scriptsize
      \begin{tabular}{lcllllllccl}
       \hline\hline
        \noalign{\smallskip}
 Object ID &    S    & \multicolumn{2}{c}{Precision}           &   F   & Precision    & Lens corr.    & Precision    &   R.A.       &     DEC    &   z$_{spec}$       \\ 
ISO\_2390\ &  (ADU)  & \multicolumn{2}{c}{  (ADU) \&}          & (mJy) & +/-mJy       &   F\,(mJy)    & +/-mJy       &              &            &    if known        \\
           &         & \multicolumn{2}{c}{Significance}        &       & (1$\sigma $) &               & (1$\sigma $) &              &            &                    \\
           &         & \multicolumn{2}{c}{[$S/\sigma_{floor}$]}&       &              &               &              &              &            &                    \\
\noalign{\smallskip} \hline         
\noalign{\smallskip}
  02 -     & 0.165   & 0.033 & [10.3]                          & 0.201 &  0.044       &    0.153      &    0.035     & 21: 53: 29.5 & 17: 42: 14 & {\it 0.700}        \\
  03 -     & 0.173   & 0.033 & [13]                            & 0.217 &  0.047       &    0.166      &    0.038     & 21: 53: 30.3 & 17: 41: 45 & {\it 0.700}        \\
  08a -    & 0.103   & 0.027 & [7.7]                           & 0.118 &  0.038       &    0.071      &    0.025     & 21: 53: 31.2 & 17: 42: 25 & 0.648$^K$          \\
  09       & 0.105   & 0.027 & [7.8]                           & 0.121 &  0.038       &    0.087      &    0.028     & 21: 53: 32.0 & 17: 41: 24 & {\it 0.700}        \\
  13       & 0.131   & 0.028 & [8.2]                           & 0.155 &  0.039       &    0.069      &    0.022     & 21: 53: 32.3 & 17: 42: 48 & {\it 0.700}        \\
  15       & 0.231   & 0.041 & [17.2]                          & 0.298 &  0.058       &    0.250      &    0.050     & 21: 53: 32.6 & 17: 41: 19 & 0.34$^{L1}$        \\
  18       & 0.29    & 0.042 & [21.6]                          & 0.382 &  0.059       &    0.052      &    0.019     & 21: 53: 33.0 & 17: 42: 09 & 2.7$^C$            \\
  24 -     & 0.275   & 0.039 & [12]                            & 0.372 &  0.051       &    0.288      &    0.045     & 21: 53: 33.8 & 17: 40: 49 & {\it 0.700}        \\
  25       & 0.295   & 0.042 & [22]                            & 0.389 &  0.059       &    0.188      &    0.044     & 21: 53: 33.8 & 17: 42: 41 & 1.467$^C$          \\
  27       & 0.247   & 0.041 & [18]                            & 0.321 &  0.058       &    0.103      &    0.031     & 21: 53: 34.2 & 17: 42: 21 & 0.913$^P$          \\
  28a -    & 0.346   & 0.041 & [26]                            & 0.461 &  0.058       &    0.051      &    0.018     & 21: 53: 34.3 & 17: 42: 02 & 0.913$^P$          \\
  33 -     & 0.14    & 0.033 & [10.5]                          & 0.170 &  0.047       &    0.102      &    0.031     & 21: 53: 35.4 & 17: 42: 36 & 0.628$^K$          \\
  42a -    & 0.116   & 0.028 & [7.2]                           & 0.135 &  0.038       &    0.033      &    0.013     & 21: 53: 38.0 & 17: 41: 56 & 0.9$^K$            \\
\noalign{\smallskip}
\noalign{\smallskip}
  22       & 0.058   & 0.016 & [4.33]                          & 0.054 &  0.023       &    0.032      &    0.014     & 21: 53: 33.6 & 17: 41: 15 & {\it 0.700}        \\
  29 -     & 0.062   & 0.016 & [4.6]                           & 0.060 &  0.023       &    0.037      &    0.015     & 21: 53: 34.4 & 17: 41: 19 & 0.53$^K$           \\
  36 -     & 0.075   & 0.019 & [5.6]                           & 0.078 &  0.027       &    0.033      &    0.013     & 21: 53: 36.6 & 17: 42: 16 & {\it 0.700}        \\
  40 -     & 0.093   & 0.016 & [5.8]                           & 0.104 &  0.021       &    0.078      &    0.017     & 21: 53: 37.5 & 17: 42: 30 & 0.42$^K$           \\
\noalign{\smallskip}
\noalign{\smallskip}
  04 -     & 0.146   & 0.056 & [2.7]                           & 0.100 &  0.084       &    0.073      &    0.062     & 21: 53: 30.4 & 17: 42: 56 & {\it 0.700}        \\
  07       & 0.254   & 0.033 & [16]                            & 0.320 &  0.044       &    0.320      &    0.044     & 21: 53: 31.2 & 17: 41: 32 & {\it 0.246$^{L1}$} \\
  19 -     & 0.074   & 0.016 & [4.6]                           & 0.078 &  0.021       &    0.043      &    0.013     & 21: 53: 33.1 & 17: 42: 49 & {\it 0.700}        \\
  23       & 0.216   & 0.056 & [4.0]                           & 0.206 &  0.084       &    0.206      &    0.084     & 21: 53: 33.7 & 17: 43: 23 & {\it 0.247$^{Y/N}$}\\
  26 -     & 0.302   & 0.069 & [5.6]                           & 0.335 &  0.104       &    0.281      &    0.088     & 21: 53: 34.2 & 17: 40: 27 & {\it 0.700}        \\
  30 -     & 0.217   & 0.056 & [4.0]                           & 0.207 &  0.084       &    0.170      &    0.070     & 21: 53: 34.4 & 17: 43: 23 & {\it 0.700}        \\
  32 -     & 0.202   & 0.056 & [3.74]                          & 0.185 &  0.084       &    0.156      &    0.071     & 21: 53: 35.2 & 17: 43: 26 & {\it 0.700}        \\
  37       & 0.391   & 0.043 & [29.2]                          & 0.524 &  0.061       &    0.524      &    0.061     & 21: 53: 36.6 & 17: 41: 44 & {\it 0.230$^{L1}$} \\
  41 -     & 0.114   & 0.024 & [5]                             & 0.160 &  0.032       &    0.160      &    0.032     & 21: 53: 37.5 & 17: 41: 09 & {\it 0.239$^{L1}$} \\
  43 -     & 0.074   & 0.016 & [4.6]                           & 0.078 &  0.021       &    0.046      &    0.014     & 21: 53: 38.3 & 17: 42: 17 & {\it 0.700}        \\
  45 -     & 0.231   & 0.056 & [4.3]                           & 0.228 &  0.084       &    0.161      &    0.061     & 21: 53: 38.9 & 17: 42: 27 & {\it 0.700}        \\
\noalign{\smallskip}
\noalign{\smallskip}
\multicolumn{10}{l}{  Redshift references :  P  = Pell\'o et al. 1991; L1 = LeBorgne et al. 1991; Y  = Yee et al. 1996; C  = Cowie et al. 2001; } \\ 
\multicolumn{10}{l}{ \ \ \ \ \ \ \ \ \ \ \ \ \ \ \ \ \ \ \ \ \ \ \ \ \ \ \ \ K  = Kneib, unpublished; N  = NED. }                             \\
\noalign{\smallskip}
\noalign{\smallskip}
    \hline                                              
   \end{tabular}
  \end{center}   
\end{table*}                                            


\vskip -10pt

\begin{table*}[!ht]
 \caption{\em 7\,$\mu$m sources detected in the field of Abell 370.
The significance of the columns is explained in the text
(Sec.\ref{sec:tables}). A horizontal blank line divides the table
into two parts. The upper section holds non-cluster sources detected
above the 80\% completeness limit.  The bottom section holds cluster
sources and any non-cluster sources not meeting statistical criteria
for inclusion in the counts. (See
Sec.\ref{sec:find_source} \& Sec.\ref{sec:monte_carlo}.)  When the redshift is 
given in italics it has been assigned to 0 for stars, or to the median
redshift of the background population, $\sim$0.7, for background objects
lacking spectroscopic redshift (see Sec.~\ref{sec:redshift} and 
\ref{sec:discuss_counts_plots}), or the
source is in the cluster. The letters associated with z values identify
the origin of the z value, expanded at the end of the table.  Source
IDs accompanied by (+) signs are sources detected at 7\,$\mu$m, but not at
15\,$\mu$m. The mean redshift of the cluster is z = 0.374.
}
  \label{tab:a370lw2tab}
   \begin{center}
    \leavevmode
     \scriptsize
      \begin{tabular}{lcllllllccl}
       \hline\hline
        \noalign{\smallskip}
 Object ID  &   S    & \multicolumn{2}{c}{Precision}            &   F   & Precision    & Lens corr. & Precision    &   R.A.      &     DEC    &   z$_{spec}$                  \\ 
 ISO\_A370\ & (ADU)  & \multicolumn{2}{c}{  (ADU) \&}           & (mJy) & +/-mJy       &  F\,(mJy)  & +/-mJy       &             &            &    if known                   \\
            &        & \multicolumn{2}{c}{Significance}         &       & (1$\sigma $) &            & (1$\sigma $) &             &            &                               \\
            &        & \multicolumn{2}{c}{[$S/\sigma_{floor}$]} &       &              &            &              &             &            &                               \\
\noalign{\smallskip}\hline
\noalign{\smallskip}
  07 +      &  0.464 & 0.033 & [14.6]                           & 0.444 &   0.033      &   0.444    &    0.033     & 2: 39: 51.1 & -1: 32: 28 & fgnd.                         \\
  09        &  0.176 & 0.033 & [5.3]                            & 0.149 &   0.034      &   0.069    &    0.020     & 2: 39: 51.9 & -1: 36: 01 & 2.803$^I$                     \\
  15        &  1.339 & 0.11  & [40.6]                           & 1.340 &   0.113      &   1.340    &    0.113     & 2: 39: 54.7 & -1: 32: 35 & 0.045$^N$                     \\
  17        &  0.479 & 0.033 & [14.5]                           & 0.460 &   0.033      &   0.256    &    0.043     & 2: 39: 56.6 & -1: 34: 29 & 1.062$^{S9/B}$                \\
  18 +      &  5.021 & 0.52  & [109]                            & 5.155 &   0.703      &   5.155    &    0.703     & 2: 39: 56.4 & -1: 31: 39 & 0.026$^N$                     \\
\noalign{\smallskip}
\noalign{\smallskip}
  06 +      &  0.842 & 0.043 & [25.5]                           & 0.830 &   0.044      &   0.830    &    0.044     & 2: 39: 49.5 & -1: 35: 14 & {\it 0 star}                  \\
  11b +     &  0.173 & 0.033 & [5.2]                            & 0.146 &   0.013      &   0.146    &    0.013     & 2: 39: 52.7 & -1: 35: 06 & {\it 0.37 (composite)$^{S8}$} \\
  11c +     &  0.144 & 0.033 & [4.4]                            & 0.117 &   0.013      &   0.117    &    0.013     & 2: 39: 53.2 & -1: 34: 59 & {\it 0.374$^M$ }              \\
  12 +      &  0.269 & 0.032 & [8.2]                            & 0.245 &   0.034      &   0.245    &    0.034     & 2: 39: 52.9 & -1: 34: 21 & {\it 0.379$^M$ }              \\
  13 +      &  0.149 & 0.033 & [4.5]                            & 0.122 &   0.013      &   0.122    &    0.013     & 2: 39: 53.5 & -1: 33: 20 & {\it cluster}                 \\
  22        &  3.437 & 0.134 & [104]                            & 3.489 &   0.137      &   3.489    &    0.137     & 2: 39: 59.2 & -1: 34: 02 & {\it 0 star}                  \\
\noalign{\smallskip}                                    
\noalign{\smallskip}
\multicolumn{10}{l}{  Redshift references :  S8 = Soucail et al. 1988; M  = Mellier et al. 1988; I  = Ivison et al. 1998; N  = NED. }             \\            
\noalign{\smallskip}
\noalign{\smallskip}
   \hline
    \end{tabular}
     \end{center}
\end{table*}

\vskip -10pt

\begin{table*}[!ht]
 \caption{\em 7\,$\mu$m sources detected in the field of Abell 2218.
The significance of the columns is explained in the text
(Sec.\ref{sec:tables}). A horizontal blank line divides the table
into two parts. The upper section holds non-cluster sources detected
above the 80\% completeness limit.  The bottom section holds cluster
sources and any non-cluster sources not meeting statistical criteria
for inclusion in the  counts. (See
Sec.\ref{sec:find_source} \& Sec.\ref{sec:monte_carlo}.) When the redshift is  given in
italics it has been assigned to 0 for stars, or to the median
redshift of the background population, $\sim$0.7, for background objects lacking spectroscopic redshift
(see Sections~\ref{sec:redshift} and \ref{sec:discuss_counts_plots}), or the source is in the cluster. The
letters associated with z values identify the origin of the z value,
expanded at the end of the table.  Source IDs accompanied by (+) signs are
sources detected at 7\,$\mu$m, but not at 15\,$\mu$m.  The mean
redshift of the cluster is z = 0.1756.
}
  \label{tab:a2218lw2tab}
   \begin{center}
    \leavevmode
     \scriptsize
      \begin{tabular}{lcllllllccl}
       \hline\hline
        \noalign{\smallskip}
 Object ID  &   S   & \multicolumn{2}{c}{Precision}            &   F    & Precision    &  Lens corr.  &  Precision    &    R.A.      &      DEC   &   z$_{spec}$          \\ 
ISO\_A2218\ & (ADU) & \multicolumn{2}{c}{  (ADU) \& }          & (mJy)  & +/-mJy       &   F\,(mJy)   &  +/-mJy       &              &            &    if known           \\
            &       & \multicolumn{2}{c}{Significance}         &        & (1$\sigma $) &              &  (1$\sigma $) &              &            &                       \\
            &       & \multicolumn{2}{c}{[$S/\sigma_{floor}$]} &        &              &              &               &              &            &                       \\
\noalign{\smallskip}\hline
\noalign{\smallskip}
  02       & 0.236  & 0.026 & [11.2]                           & 0.220  &   0.027      &   0.189      &    0.020      & 16: 35: 32.2 & 66: 12: 04 & 0.53$^K$              \\
  17       & 0.11   & 0.02  & [6.1]                            & 0.099  &   0.021      &   0.078      &    0.017      & 16: 35: 43.6 & 66: 11: 57 & {\it 0.700}           \\ 
  27       & 0.451  & 0.04  & [25]                             & 0.451  &   0.042      &   0.451      &    0.042      & 16: 35: 48.9 & 66: 13: 02 & 0.103$^{L2}$          \\
  35       & 0.111  & 0.02  & [6.2]                            & 0.100  &   0.021      &   0.039      &    0.012      & 16: 35: 53.5 & 66: 12: 58 & 0.474$^{Eb}$          \\
  38       & 0.124  & 0.018 & [6.9]                            & 0.113  &   0.019      &   0.027      &    0.006      & 16: 35: 55.1 & 66: 11: 51 & 1.034$^P$             \\
  42       & 0.09   & 0.022 & [5]                              & 0.078  &   0.023      &   0.049      &    0.018      & 16: 35: 55.7 & 66: 13: 16 & 0.45$^K$              \\
  51       & 0.137  & 0.018 & [7.6]                            & 0.127  &   0.019      &   0.065      &    0.013      & 16: 35: 58.1 & 66: 11: 13 & {\it 0.700}           \\
  52       & 0.112  & 0.02  & [6.2]                            & 0.101  &   0.021      &   0.039      &    0.012      & 16: 35: 59.0 & 66: 11: 48 & 0.55$^K$              \\
  61a      & 0.133  & 0.021 & [7.4]                            & 0.128  &   0.022      &   0.094      &    0.022      & 16: 36: 03.8 & 66: 13: 33 & {\it 0.700}           \\ 
  67       & 0.129  & 0.018 & [7.2]                            & 0.119  &   0.019      &   0.086      &    0.018      & 16: 36: 06.4 & 66: 12: 30 & {\it 0.700}           \\
  68       & 0.094  & 0.031 & [5.2]                            & 0.088  &   0.032      &   0.074      &    0.024      & 16: 36: 07.5 & 66: 13: 02 & 0.42$^K$              \\
\noalign{\smallskip}
\noalign{\smallskip}
  03  +    & 0.128  & 0.038 & [3.3]                            & 0.090  &   0.036      &   0.090      &    0.036      & 16: 35: 33.1 & 66: 12: 58 & {\it 0 star}          \\
  04  +    & 0.907  & 0.22  & [13]                             & 0.818  &   0.124      &   0.818      &    0.124      & 16: 35: 33.7 & 66: 11: 12 & {\it 0 star}          \\
  06       & 0.262  & 0.026 & [12.5]                           & 0.247  &   0.027      &   0.247      &    0.027      & 16: 35: 34.6 & 66: 12: 34 & {\it 0.180$^{L2}$}    \\
  11  +    & 0.191  & 0.022 & [9.1]                            & 0.173  &   0.023      &   0.173      &    0.023      & 16: 35: 40.0 & 66: 13: 21 & {\it 0 star}          \\
  12  +    & 0.128  & 0.069 & [1.9]                            & 0.081  &   0.066      &   0.081      &    0.066      & 16: 35: 40.4 & 66: 14: 25 & {\it 0 star}          \\
  14  +    & 0.172  & 0.029 & [8.2]                            & 0.153  &   0.030      &   0.153      &    0.030      & 16: 35: 41.3 & 66: 13: 47 & {\it 0.183$^{L2}$}    \\
  15b +    & 0.431  & 0.04  & [24]                             & 0.431  &   0.042      &   0.431      &    0.042      & 16: 35: 42.4 & 66: 12: 10 & {\it 0 star}          \\
  16  +    & 0.133  & 0.069 & [1.9]                            & 0.085  &   0.066      &   0.085      &    0.066      & 16: 35: 42.8 & 66: 14: 28 & {\it 0 star}          \\
  18  +    & 0.131  & 0.018 & [7.3]                            & 0.121  &   0.019      &   0.121      &    0.019      & 16: 35: 43.9 & 66: 13: 21 & {\it 0.178$^{L2}$}    \\
  21  +    & 0.079  & 0.027 & [4.4]                            & 0.073  &   0.028      &   0.073      &    0.028      & 16: 35: 46.7 & 66: 12: 22 & {\it 0.1638$^{S01}$}  \\
  22  +    & 0.041  & 0.018 & [2.3]                            & 0.034  &   0.027      &   0.034      &    0.027      & 16: 35: 47.1 & 66: 13: 16 & {\it 0.18$^{L2}$}     \\
  23  +    & 0.143  & 0.038 & [3.7]                            & 0.104  &   0.036      &   0.104      &    0.036      & 16: 35: 47.3 & 66: 14: 45 & {\it 0.1545$^N$ }     \\
  24  +    & 0.178  & 0.022 & [9.9]                            & 0.169  &   0.023      &   0.169      &    0.023      & 16: 35: 47.6 & 66: 11: 08 & {\it 0.164$^{L2}$}    \\
  28  +    & 0.243  & 0.037 & [13.5]                           & 0.236  &   0.039      &   0.236      &    0.039      & 16: 35: 48.9 & 66: 12: 44 & {\it 0.172$^{L2}$}    \\
  30       & 0.245  & 0.037 & [13.6]                           & 0.238  &   0.039      &   0.238      &    0.039      & 16: 35: 50.2 & 66: 12: 24 & {\it 0.178$^{L2}$}    \\
  32  +    & 0.07   & 0.018 & [3.9]                            & 0.063  &   0.018      &   0.063      &    0.018      & 16: 35: 51.3 & 66: 13: 13 & {\it 0.1537$^N$}      \\
  33  +    & 0.155  & 0.02  & [8.6]                            & 0.145  &   0.021      &   0.145      &    0.021      & 16: 35: 51.6 & 66: 12: 34 & {\it 0.164$^{L2}$}    \\
  36b +    & 0.107  & 0.02  & [6]                              & 0.096  &   0.021      &   0.096      &    0.021      & 16: 35: 54.6 & 66: 14: 01 & {\it 0.1598$^N$}      \\
  39  +    & 0.131  & 0.031 & [6.2]                            & 0.111  &   0.032      &   0.111      &    0.032      & 16: 35: 54.9 & 66: 10: 15 & {\it 0.1517$^N$}      \\
  40  +    & 0.031  & 0.018 & [1.7]                            & 0.023  &   0.027      &   0.005      &    0.005      & 16: 35: 55.0 & 66: 12: 38 & {\it 0.702$^{SED}$}   \\
  41  +    & 0.085  & 0.027 & [4.7]                            & 0.079  &   0.028      &   0.079      &    0.028      & 16: 35: 55.3 & 66: 14: 14 & {\it cluster}         \\
  45  +    & 0.227  & 0.037 & [12.6]                           & 0.220  &   0.039      &   0.220      &    0.039      & 16: 35: 56.5 & 66: 11: 54 & {\it 0.1768$^{L2}$}   \\
  47  +    & 0.184  & 0.037 & [10.2]                           & 0.175  &   0.039      &   0.175      &    0.039      & 16: 35: 57.1 & 66: 11: 09 & {\it 0.176$^{L2}$}    \\
  48       & 0.101  & 0.038 & [2.6]                            & 0.065  &   0.036      &   0.056      &    0.030      & 16: 35: 57.4 & 66: 14: 38 & {\it 0.700}           \\
  49       & 0.135  & 0.031 & [6.4]                            & 0.115  &   0.032      &   0.115      &    0.032      & 16: 35: 58.0 & 66: 14: 23 & {\it 0.15$^{SED}$}    \\ 
  54  +    & 0.111  & 0.02  & [6.2]                            & 0.100  &   0.021      &   0.100      &    0.021      & 16: 35: 59.2 & 66: 12: 06 & {\it 0.180$^{L2}$}    \\
  56       & 0.454  & 0.04  & [25]                             & 0.454  &   0.042      &   0.454      &    0.042      & 16: 36: 00.7 & 66: 13: 21 & {\it 0 star}          \\
  57  +    & 0.198  & 0.038 & [11]                             & 0.190  &   0.040      &   0.190      &    0.040      & 16: 36: 02.1 & 66: 12: 34 & {\it 0.175$^{L2}$}    \\
  58  +    & 0.142  & 0.02  & [7.9]                            & 0.132  &   0.021      &   0.132      &    0.021      & 16: 36: 02.3 & 66: 11: 53 & {\it 0.183$^{L2}$}    \\
  60       & 0.053  & 0.026 & [3]                              & 0.046  &   0.027      &   0.034      &    0.021      & 16: 36: 04.1 & 66: 13: 03 & 0.2913$^{L2}$         \\
  61b +    & 0.166  & 0.022 & [9.2]                            & 0.158  &   0.023      &   0.158      &    0.023      & 16: 36: 04.1 & 66: 13: 27 & {\it 0.174$^{L2}$}    \\
  62  +    & 0.142  & 0.02  & [7.9]                            & 0.132  &   0.021      &   0.132      &    0.021      & 16: 36: 03.9 & 66: 11: 41 & {\it 0.176$^{L2}$}    \\
  64  +    & 0.192  & 0.022 & [9.1]                            & 0.174  &   0.023      &   0.174      &    0.023      & 16: 36: 04.4 & 66: 10: 56 & {\it cluster}         \\
  66  +    & 0.097  & 0.02  & [5.4]                            & 0.085  &   0.023      &   0.085      &    0.023      & 16: 36: 06.3 & 66: 12: 48 & {\it 0.168$^{L2}$}    \\
  71  +    & 0.272  & 0.05  & [7]                              & 0.220  &   0.043      &   0.220      &    0.043      & 16: 36: 09.2 & 66: 11: 18 & {\it 0 star}          \\
  72       & 0.921  & 0.037 & [44]                             & 0.934  &   0.039      &   0.934      &    0.039      & 16: 36: 09.6 & 66: 11: 57 & {\it 0 star}          \\
  75  +    & 0.102  & 0.038 & [2.6]                            & 0.066  &   0.036      &   0.066      &    0.036      & 16: 36: 11.7 & 66: 12: 52 & {\it cluster}         \\
\noalign{\smallskip}                                    
\noalign{\smallskip}
\multicolumn{10}{l}{  Redshift references :  P = Pell\'o et al. 1991; L2 = LeBorgne et al. 1992; Eb = Ebbels et al. 1998; S01 = Smail et al. 2001;  }                       \\
\multicolumn{10}{l}{ \ \ \ \ \ \ \ \ \ \ \ \ \ \ \ \ \ \ \ \ \ \ \ \ \ \ \ \ K = Kneib, unpublished; N   = NED; SED = spectral fitting.}                                  \\
\noalign{\smallskip}
\multicolumn{10}{l}{  {\it ISO\_A2218\_28+ and 45+ are extended sources even at ISO resolution. The reduction algorithm is not adapted to treat}}                         \\
\multicolumn{10}{l}{  {\it extended structure and the transformation of extended structure has not been characterised. No attempt has been made}  }                       \\
\multicolumn{10}{l}{  {\it here to correct for this in the case of these two cluster sources. For these sources, the quoted fluxes should be regarded } }                 \\
\multicolumn{10}{l}{  {\it as probable underestimates, especially in the case of 28+.}  } \\
\noalign{\smallskip}
\noalign{\smallskip}
   \hline                                                        
      \end{tabular}                                              
   \end{center}                                                  
\end{table*}

\vskip -10pt

\begin{table*}[!ht]
 \caption{\em 7\,$\mu$m sources detected in the field of Abell 2390.
The significance of the columns is explained in the text
(Sec.\ref{sec:tables}). A horizontal blank line divides the table into two parts. The top
section holds non-cluster sources detected above the 80\% completeness limit.  The bottom section holds
cluster sources and any non-cluster sources not meeting statistical criteria for inclusion in the 
counts. (See
Sec.\ref{sec:find_source} \& Sec.\ref{sec:monte_carlo}.) When the redshift is 
given in italics it
has been assigned to 0 for stars, or to the median redshift ot the background population,
$\sim$0.7, for background objects lacking spectroscopic redshift (see
Sections~\ref{sec:redshift} and \ref{sec:discuss_counts_plots}), or the source is in the cluster. The letters associated with z values
identify the origin of the z value, expanded at the end of the table. 
Source IDs accompanied by (+) signs are sources detected at 7\,$\mu$m, but
not at 15\,$\mu$m. The mean redshift of the cluster is z = 0.232.
}
  \label{tab:a2390lw2tab}
   \begin{center}
    \leavevmode
     \scriptsize
      \begin{tabular}{lcllllllccl}
       \hline\hline
        \noalign{\smallskip}
 Object ID    &   S    & \multicolumn{2}{c}{Precision}             &   F   & Precision    & Lens corr.    & Precision    &   R.A.        &     DEC    &   z$_{spec}$        \\ 
 ISO\_A2390\  & (ADU)  & \multicolumn{2}{c}{  (ADU) \&}            & (mJy) & +/-mJy       &    F\,(mJy)   & +/-mJy       &               &            &    if known         \\
              &        & \multicolumn{2}{c}{Significance}          &       & (1$\sigma $) &               & (1$\sigma $) &               &            &                     \\
              &        & \multicolumn{2}{c}{[$S/\sigma_{floor}$]}  &       &              &               &              &               &            &                     \\
\noalign{\smallskip}\hline
\noalign{\smallskip}
  15         & 0.125   & 0.017 & [8.9]                             & 0.118 &   0.021      &     0.100     &    0.019     & 21: 53: 32.7  & 17: 41: 22 &  0.34 $^K$          \\
  18         & 0.11    & 0.017 & [7.9]                             & 0.103 &   0.030      &     0.015     &    0.006     & 21: 53: 33.1  & 17: 42: 12 &  2.7$^C$            \\
  22         & 0.087   & 0.017 & [6.2]                             & 0.079 &   0.029      &     0.046     &    0.018     & 21: 53: 33.8  & 17: 41: 17 &  {\it 0.700}        \\
  25         & 0.19    & 0.017 & [13.6]                            & 0.186 &   0.021      &     0.091     &    0.019     & 21: 53: 34.0  & 17: 42: 43 &  1.467$^C$          \\
  27         & 0.158   & 0.017 & [11.3]                            & 0.153 &   0.021      &     0.051     &    0.014     & 21: 53: 34.4  & 17: 42: 23 &  0.913$^P$          \\
\noalign{\smallskip}                         
\noalign{\smallskip}                                 
  06  +      & 0.051   & 0.017 & [3.2]                             & 0.028 &   0.018      &     0.028     &    0.018     & 21: 53: 31.1  & 17: 41: 23 &  {\it star$^{L1}$}  \\
  07         & 0.151   & 0.017 & [10.8]                            & 0.243 &   0.021      &     0.243     &    0.021     & 21: 53: 31.3  & 17: 41: 35 &  {\it 0.246$^{L1}$} \\
  08b +      & 0.094   & 0.017 & [6.7]                             & 0.086 &   0.029      &     0.086     &    0.029     & 21: 53: 31.4  & 17: 42: 30 &  {\it 0.2205$^{L1}$}\\ 
  09         & 0.046   & 0.017 & [3.3]                             & 0.037 &   0.011      &     0.025     &    0.008     & 21: 53: 32.3  & 17: 41: 30 &  {\it 0.700}        \\
  10  +      & 0.104   & 0.017 & [7.4]                             & 0.097 &   0.030      &     0.097     &    0.030     & 21: 53: 32.2  & 17: 41: 43 &  {\it 0 star}       \\
  13         & 0.071   & 0.017 & [4.4]                             & 0.049 &   0.018      &     0.049     &    0.008     & 21: 53: 32.4  & 17: 42: 52 &  {\it 0.700}        \\ 
  20  +      & 0.086   & 0.017 & [6.1]                             & 0.078 &   0.029      &     0.078     &    0.029     & 21: 53: 33.4  & 17: 41: 59 &  {\it 0.23 $^{L1}$} \\
  21  +      & 0.082   & 0.017 & [5.9]                             & 0.074 &   0.029      &     0.074     &    0.029     & 21: 53: 33.6  & 17: 41: 29 &  {\it 0.218$^{L1}$} \\
  23         & 0.147   & 0.019 & [2.7]                             & 0.091 &   0.064      &     0.091     &    0.064     & 21: 53: 33.8  & 17: 43: 24 &  {\it 0.247$^{Y/N}$}\\
  28b +      & 0.095   & 0.017 & [6.8]                             & 0.087 &   0.030      &     0.087     &    0.030     & 21: 53: 34.4  & 17: 42: 00 &  {\it 0.23 $^{L1}$} \\
  31  +      & 0.062   & 0.017 & [4.4]                             & 0.053 &   0.026      &     0.053     &    0.026     & 21: 53: 35.1  & 17: 41: 54 &  {\it 0.23 $^{L1}$} \\
  34  +      & 0.059   & 0.017 & [3.7]                             & 0.037 &   0.018      &     0.037     &    0.018     & 21: 53: 35.4  & 17: 41: 11 &  {\it 0.209$^{L1}$} \\
  35  +      & 0.161   & 0.017 & [10.1]                            & 0.143 &   0.029      &     0.143     &    0.029     & 21: 53: 36.2  & 17: 41: 13 &  {\it 0.249$^{L1}$} \\
  37         & 0.359   & 0.017 & [25.6]                            & 0.361 &   0.026      &     0.361     &    0.026     & 21: 53: 36.8  & 17: 41: 46 &  {\it 0.23 $^{L1}$} \\
  38  +      & 0.118   & 0.017 & [7.4]                             & 0.098 &   0.017      &     0.098     &    0.017     & 21: 53: 37.4  & 17: 41: 24 &  {\it 0.229$^{L1}$} \\
  39  +      & 0.086   & 0.017 & [5.4]                             & 0.064 &   0.018      &     0.064     &    0.018     & 21: 53: 37.4  & 17: 41: 45 &  {\it 0.23 $^{L1}$} \\
  42b +      & 0.067   & 0.017 & [4.2]                             & 0.045 &   0.018      &     0.045     &    0.018     & 21: 53: 37.8  & 17: 41: 55 &  {\it 0.236$^{L1}$} \\
  44  +      & 0.166   & 0.017 & [10.4]                            & 0.148 &   0.029      &     0.148     &    0.029     & 21: 53: 38.3  & 17: 41: 47 &  {\it 0.233$^{L1}$} \\
  47  +      & 0.185   & 0.019 & [3.4]                             & 0.135 &   0.077      &     0.135     &    0.077     & 21: 53: 39.8  & 17: 41: 56 &  {\it 0 star$^{L1}$}\\
  49  +      & 0.138   & 0.017 & [2.5]                             & 0.081 &   0.064      &     0.051     &    0.042     & 21: 53: 38.3  & 17: 41: 47 &  {\it 0.700}        \\
\noalign{\smallskip}
\noalign{\smallskip}
\multicolumn{10}{l}{  Redshift references :  P = Pell\'o et al. 1991; L1 = LeBorgne et al. 1991; C  = Cowie et al. 2001; Y  = Yee et al. 1996; }\\
\multicolumn{10}{l}{ \ \ \ \ \ \ \ \ \ \ \ \ \ \ \ \ \ \ \ \ \ \ \ \ \ \ \ \ K  = Kneib, unpublished; N  = NED. }                                  \\
\noalign{\smallskip}
\noalign{\smallskip}
   \hline
      \end{tabular}
   \end{center}
\end{table*}

\vskip 75pt   

\section{Results - Number counts and source properties}
\label{sec:counts}

We have extracted 15\,$\mu$m and 7\,$\mu$m source counts for the three
clusters to 80\% and (for 15\,$\mu$m only) 50\% completeness levels,
which approximate to 5$\sigma$ and 4$\sigma$ minimum thresholds,
respectively. At these significance levels Eddington bias (Eddington
1913, Hogg \& Turner 1998) is not expected to affect the
counts, because all but a very few of the
counted sources are above 5$\sigma$ (refer to Tables~\ref{tab:a370lw3tab}
to \ref{tab:a2390lw2tab}, and note that, as explained in Sec.\ref{sec:tables}, 
the sources in the bottom section of each table have not been used for source
counting.)  The two faintest counted 15\,$\mu$m sources have
significance levels of 4.3$\sigma$
and 4.6$\sigma$ respectively, all others being above 5$\sigma$.  It is 
worth noting that, because of the variation of sensitivity and
lensing amplification over the fields, sources detected with a high
apparent flux level may appear at the faint flux end of the counts
plot, which refers to intrinsic (lensing-corrected) flux,
and vice versa. It follows that sources of all accepted significance levels are
spread throughout the range of intrinsic flux covered in the counts.
This further mitigates concern about any bias selectively affecting the faint
end of the counts and adds significance to the agreement between the counts to 80\%
and 50\% completeness.

Table~\ref{tab:numbers} lists the number of 15\,$\mu$m and 7\,$\mu$m
non-stellar, non-cluster, field sources falling above the 80\% and 
(for 15\,$\mu$m only) above the 50\% completeness thresholds\footnote {Note that the 
number of sources for
which photometry is quoted in the main tables is greater than in Table~\ref{tab:numbers} 
because the signal thresholds for inclusion in the photometry lists are
lower than the thresholds for counting, and also because the photometry
tables quote the fluxes for cluster galaxies and stars as well as for
field galaxies.}. 

At 15\,$\mu$m the flux interval bins for count plotting were chosen in such 
a way as to have at least 9 counts per bin for the 50\% completeness counts
while maintaining reasonable bin widths across the flux range. At 7\,$\mu$m
 we accepted 7 to 8 counts per bin and restricted to 80\% completeness counts (in large 
part because the lens properties in the flux-density region between 50\% and
80\% completeness at 7\,$\mu$m led to only a small increase in survey area when
passing from 80\% to 50\% completeness, and so there was little to be gained
by extracting 50\% completeness counts at 7\,$\mu$m.)

Table~\ref{tab:binnumbers} lists the number of sources detected in each
of the count bins appearing in the counts plots (Figs.\ref{countslw3_diff}, 
~\ref{countslw3_integ}, ~\ref{countslw2_diff} and
~\ref{countslw2_integ}, ~\ref{lari_mod}).

\subsection{Corrections applied in deriving the counts}
\label{sec:area_counts}

Several factors were taken into account in deriving source counts from the
raw list of detected sources (Tables~\ref{tab:a370lw3tab}
to \ref{tab:a2390lw2tab}). These factors are listed here, and are
discussed throughout this paper in the sections indicated :

\begin{itemize}
  \item lensing amplification and area dilation (see Sec.\ref{sec:lens_correct})
  \item completeness of the counts (see Sec.\ref{sec:monte_carlo}
  \& \ref{sec:lens_correct})
  \item cluster contamination (see Sec.\ref{sec:redshift})
  \item appropriate selection of the flux value at which to plot the
  counts for each bin (see Sec.\ref{sec:plot_where}).
\end{itemize}

\subsubsection{Removal of cluster contamination from the galaxy counts}
\label{sec:redshift}

In order to determine the number density of field galaxies it is
necessary to eliminate cluster galaxies from the counts.  In cases
where redshift information is available this is straightforward.
However, we have redshifts for 89 of the 145 sources detected. 
Of the remaining 56 sources, 34 meet the S/N and completeness 
criteria for inclusion in the source counts. 

A strong correlation exists between the 15 to 7\,$\mu$m colour of
the 89 sources with known redshift and the galaxy status as a cluster
or non-cluster-member. This correlation enables sorting of the 34
unknown-redshift countable sources into cluster and non-cluster categories,
so that they could be included or excluded from the source counts as 
appropriate. 

The following approach was adopted to perform the sorting.  The sources from 
the 15\,$\mu$m sample having known
redshift were sorted into the categories foreground, background and
cluster-member. This was done independently for sources having
15\,$\mu$m-only detections, and for sources seen at both 15 and
7\,$\mu$m.  

\begin{figure*}
\centering
\includegraphics[angle=0,width=0.85\textwidth,bb=17 220 592 773]{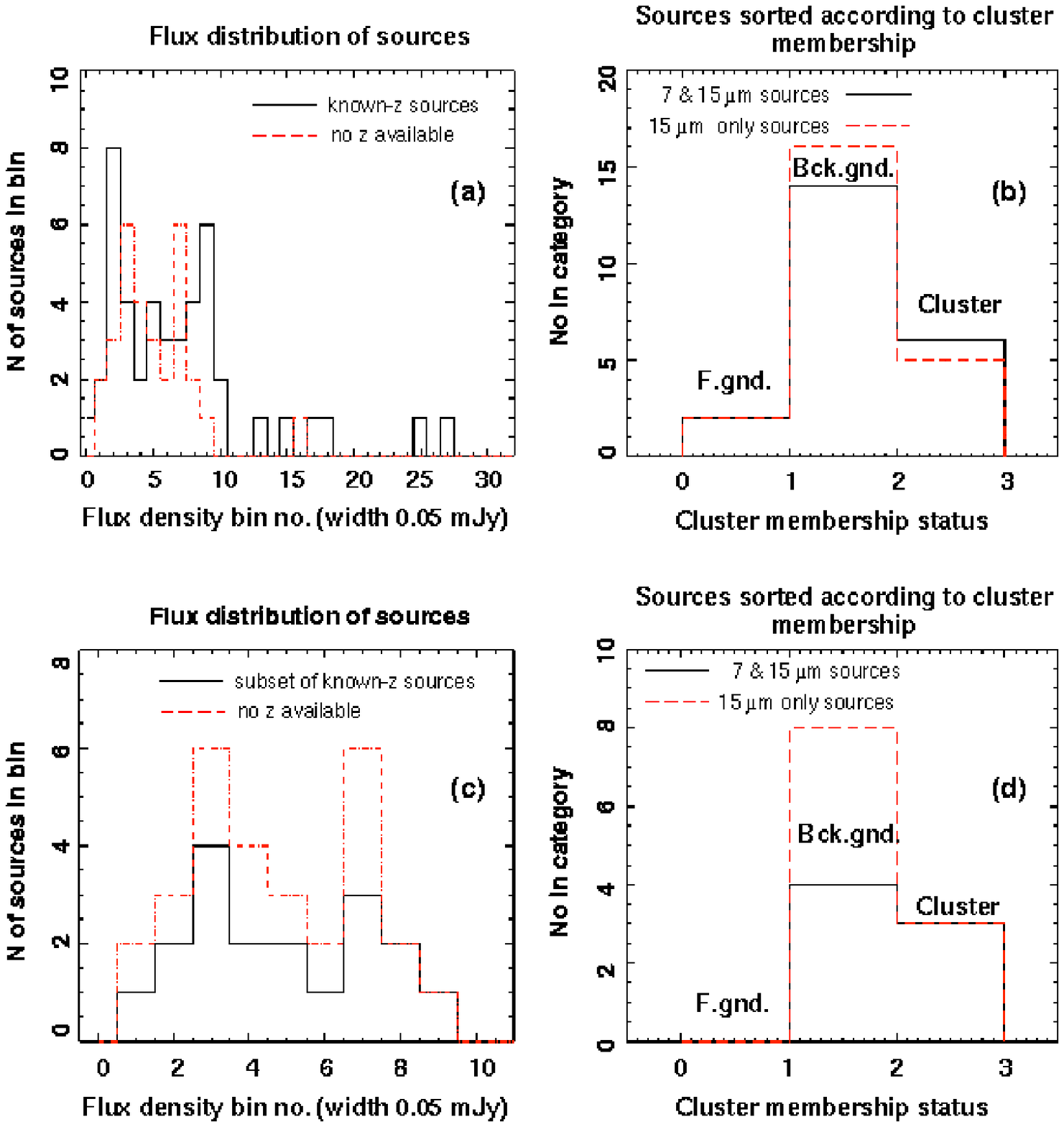}
\caption[]{(a) gives, for the 15$\mu$m sample,
 the distribution with respect to flux-density of all sources having
 available redshift information (solid line), superimposed on the
 distribution of countable sources lacking redshift information (broken
 line). The known-z sample has bright outliers relative to the
 unknown-z  sample. Nevertheless, (b) shows that between 73 and 78\% of
 the galaxies seen at both 7 and  15\,$\mu$m (solid line) or at
 15\,$\mu$m-only (broken line) in this sample are non-cluster galaxies
 (see text for further explanation).  (c) gives the distributions with
 respect to flux-density of a subset of sources in the 15$\mu$m sample
 having available redshift (solid line) and yielding a flux
 distribution closer to the distribution of countable sources lacking
 redshift information (broken line).  (d) shows that 75\% of the
 galaxies seen at 15\,$\mu$m-only (broken line) or 60\% of galaxies
 seen at both 7 and  15\,$\mu$m (solid line), in this sample, are
 non-cluster galaxies.   
} 
\label{fluxdist_rule_15}
\end{figure*}

It was found (Fig.~\ref{fluxdist_rule_15}) that
between 73\% and 78\% of the set of sources fulfilling the condition {\it [detected at
both 15 and 7\,$\mu$m (solid line) OR detected only at 15\,$\mu$m
(dashed line)]}, are non-cluster members. However, as shown in
Fig.~\ref{fluxdist_rule_15}.a,  the distribution with flux of the
known-z sources (solid line) does not quite  correspond to the flux
distribution of the unknown-z sources (broken-line), in the sense that the known-z
set has bright outliers.   So the sorting
rule defined on the known-z set may not apply well to the unknown-z set.

We therefore selected from the set of sources with known z, a subset
matching in its flux distribution (Fig.~\ref{fluxdist_rule_15}.c -
solid line) the flux-distribution of the sources lacking z and
needing to be  sorted (Fig.~\ref{fluxdist_rule_15}.c - broken
line).   For that subset of known-z sources, the sorting rule
represented by Fig.~\ref{fluxdist_rule_15}.d was found.  Consistent
with the result on the full set of known-z sources, of cases detected
only at 15\,$\mu$m (dashed line) 75\% are background galaxies.  Of
known-z source detected at both  15 and 7\,$\mu$m (solid line) about
60\% are background sources.

We used these rough but adequate rules to sort the unknown-z sources
as follows: the 26 countable 15\,$\mu$m-only sources were assigned to
the category of background sources, counting them  with a weight of
0.75.  And we assigned the 8 cases of unknown-z countable sources
having both 7 and 15\,$\mu$m detections to the background category,
counting them with a weight of 0.6.  Even if we assumed, for example,
that the 0.75 weighting factor, which acts on 26 sources, were off
by, say, 0.15 (which is outside the range of any fluctuation we
obtained in different sortings on the known-z sources), the 
cumulative error for the 26 sources would be four counts distributed
across  the 5 count bins: essentially neglectable compared with other
error sources (e.g. the Poisson error on the counts themselves). It
would not significantly impact the results of this paper.\\

\vskip -8pt

\begin{figure*}
\centering
\includegraphics[angle=0,width=0.85\textwidth,bb=17 220 592 773]{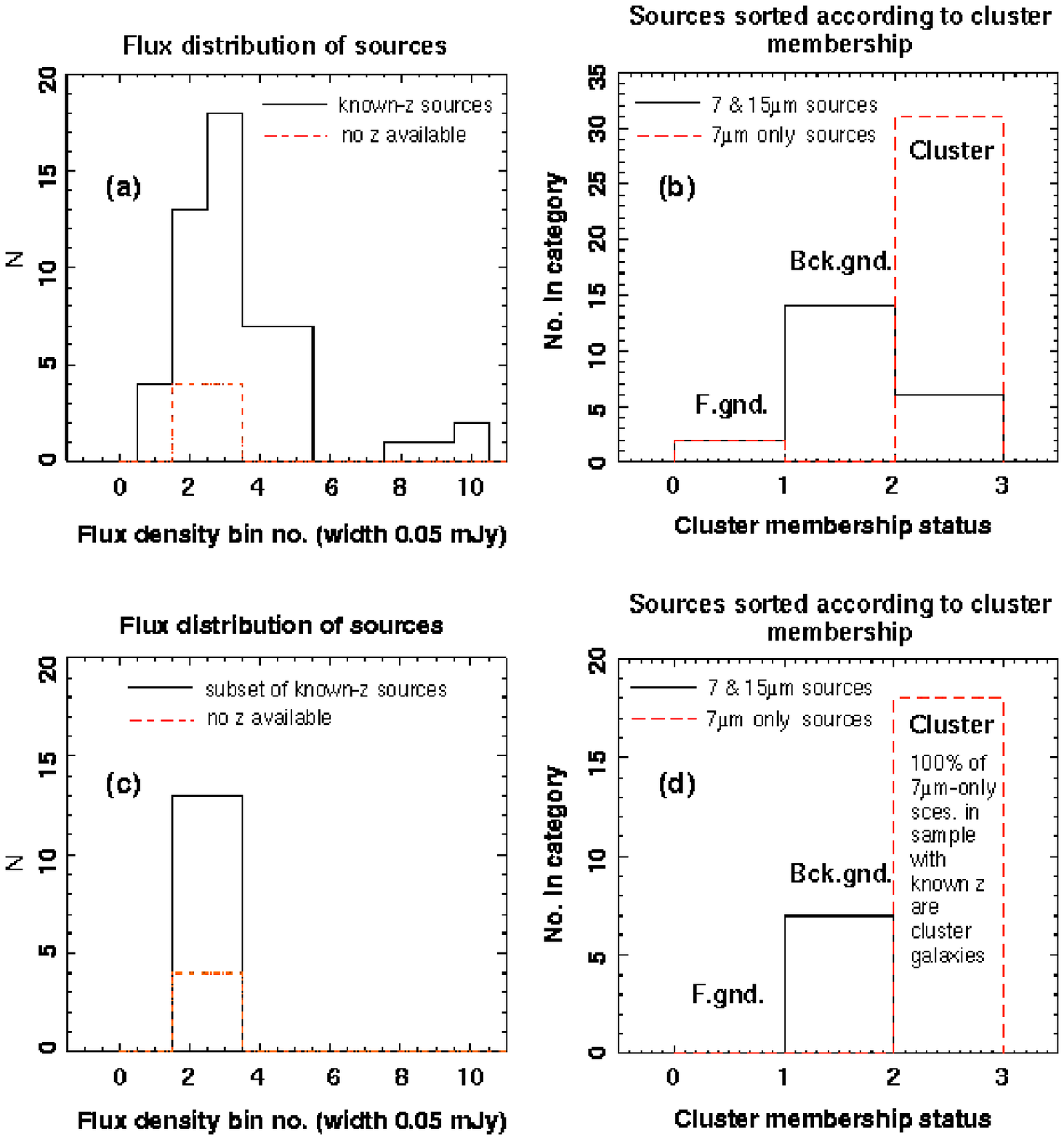}
\caption[]{(a) gives, for the 7$\mu$m sample,
 the distribution with respect to flux-density of all sources having
 available redshift information (solid line), superimposed on the
 distribution of countable sources lacking redshift information (broken
 line). The known-z sample has bright outliers relative to the
 unknown-z  sample. Nevertheless, (b) shows that 95\% of the
 galaxies seen at 7\,$\mu$m-only (broken line) are cluster members,
 while only 27\% of galaxies seen also at 15\,$\mu$m are in the
 clusters.  (c) gives the distributions with respect to flux-density
 of a subset of sources  in the 7$\mu$m sample having available
 redshift (solid line) and yielding a flux distribution closer to the
 distribution of countable sources lacking redshift information (broken
 line).  (d) shows that 100\% of the galaxies seen at 7\,$\mu$m-only
 (broken line) are cluster members,  while 100\% of galaxies seen at both
 7 and 15\,$\mu$m (solid line), in this sample,  are non-cluster
 galaxies.}
\label{fluxdist_rule_7}
\end{figure*}

Fig.~\ref{fluxdist_rule_7} describes the corresponding approach applied
to the set of 7\,$\mu$m sources.
Fig.~\ref{fluxdist_rule_7}.a  compares the flux distribution of known-z
sources having a 7-micron detection (solid line) with the flux
distribution of unknown-z otherwise countable 7\,$\mu$m sources (broken
line). Fig.~\ref{fluxdist_rule_7}.c is the corresponding plot  obtained
when the chosen set of known-z sources is matched in its flux
distribution to the unknown-z set.  Fig.~\ref{fluxdist_rule_7}.d is the
sorting rule defined on this subset of known-z sources.  It is seen
that, for sources detected only at 7\,$\mu$m (broken line) 95 to 100\% are
cluster sources, while for sources detected at both 7\,$\mu$m and 15\,$\mu$m
something between 70\% and 100\% are background sources, and we have
taken a value of 85\%.

\vskip 6pt

We therefore used these rough but adequate rules to assign the 4
unknown--z otherwise--countable 7\,$\mu$m-only sources to the category of
cluster sources, eliminating them from the counts. And we assigned the 5
cases of unknown--z otherwise--countable sources having both 7 and 15\,$\mu$m
detections to the background category, counting them with a weight of
0.85 in that category. Even if it is supposed that the weighting factor of 0.7
implied by the full population of Fig.~\ref{fluxdist_rule_7} should be
applied, the cumulative delta-contribution from the 5 sources would be
0.75 counts, distributed across all bins, which is negligible with respect 
to the error bars.

\vskip 6pt

Similar tests on individual clusters showed these sorting rules to be
robust for the survey, within about 0.l of the quoted weighting factors.

\subsubsection{Appropriate selection of the flux value at which to plot the counts for each bin}
\label{sec:plot_where}

\vskip 4pt

In order to plot the number counts reliably it is necessary to take
account of the fact that the bins are quite broad and the
value of the population density being sampled changes substantially
across a bin. A continuous function is being sampled with a crude
histogram, and so simply plotting each sample at the central
flux-density of each bin results in plotted values which may fall
well off the curve of the underlying population being sampled.\\

Assume that the population being sampled, expressed in terms of number
of sources per square degree per mJy, n, is described by a power-law
of the form $n = C.S^{a}$, S being the flux-density and C a constant.
We sample this population by counting the number of sources dN within
some flux range $S_{1}$ to $S_{2}$, and the resulting
differential counts are plotted.  We require that the data point for a value of
the differential counts dN/dS be plotted at a flux value such that the
point falls on the curve of the underlying population. 

\noindent That is :
\\

\vspace*{-0.15cm}
$\int_{{\Delta}S}n.dS = dN = \int_{{\Delta}S}C.S^{a}.dS = [C.S^{(a+1)}/(a+1)]_{S_1}^{S_2}$
\\

\vspace*{-0.15cm}
\noindent and the differential count dN/dS should be plotted at flux-density $S_{plot}$ such 
that :
\\

\vspace*{-0.3cm}
$dN/dS = $\\
$\hspace*{+1cm} [C/(a+1)].[S_{2}^{(a+1)} - S_{1}^{(a+1)}]/(S_{2}-S_{1}) = C.S_{plot}^{a}$ \\

\vspace*{-0.3cm}
\noindent which immediately gives the appropriate value of $S_{plot}$ for each
value of the differential counts, provided one has a first-order
estimate of the characteristic population index, a.

To make the differential counts plots 
we have derived a first-order estimate of the index by plotting the
counts without any weighting, deriving a best fit slope, and then
iterating. The count indices converge
to the values given on the counts plots (Figs.\ref{countslw3_diff}, \ref{countslw3_integ},
\ref{countslw2_diff} and \ref{countslw2_integ}).
[Obviously, in the case of the integral source counts plots showing sources per
square degree down to a series of flux-density limits, the data points
are always plotted at the lower end of the flux bin being integrated.]

The above approach results in Integrated Galaxy Light (IGL) totals from differential and integral
counts which are mutually consistent to within a few percent.  The IBL
values so derived also agree  well with a simple summation
$\sum_{all bins}(counts per bin \times S_{plot})$.   If the counts per
bin are not plotted as described above to take account of bias due to
the crude binning of the underlying population function, then the
different  methods of computing the background integral yield
inconsistent results at the level of at least 20\%.

\subsubsection{Propagation of errors - estimation of precision}
\label{sec:errors_of_counts}

For source photometry, the flux-density uncertainty associated with each
source is determined from the Monte Carlo simulations, and is the
observed 1-sigma scatter in recovered flux for fake-sources of
corresponding brightness inserted into the raw data cube.  The scatter
in recovered signal is not independent of the signal level, reflecting
detailed behaviour of the reduction algorithm traced by the
simulations.  The noise floor for each redundancy zone of each map is 
determined by taking the scatter of recovered signal for the faintest
fake sources recovered with at least 50\% completeness.  This is the
noise level used to determine the significance of each detection.


We fold in quadrature the Poisson error on the number of sources
detected in any given bin with a 10\% uncertainty for the value of the 
lensing model area correction and a 10\% uncertainty derived from 
fluctuations in the simulation completeness correction to arrive at the
error estimate for counts in each counting bin.

\subsubsection{Variation of area surveyed with flux - implications for cosmic variance}
\label{sec:cosmic_var}

As described in association with Fig.~\ref{how_deep1}, the area surveyed decreases 
as flux decreases, because the raster observation sensitivity is
dependent upon position (redundancy) within the raster, and also because the
amplification of the lens is dependent upon position on the sky. The
mean area surveyed for the deepest bin was
2 square arcminutes in the lensing-corrected source plane, an area 
corresponding to a mean lensing amplification of about three and an
apparent surveyed area on the sky of about 6 square arcminutes. The most
sensitive 15\,$\mu$m counts bin, plotted in the differential counts at 47\,$\mu$Jy, 
relies entirely on sources from the deepest field, A2390.  The second faintest
15\,$\mu$m  bin, plotted in the differential counts at 107\,$\mu$Jy, addresses an average
area of 8.2 square  arcminutes in the lensing-corrected source plane
and is based upon sources from  the A2390 and A2218 fields.  The
faintest 7\,$\mu$m bin reported here derives from an average
area of 8.7 square arcminutes in the lensing-corrected source plane,
with about equal contributions from the A2390 and A2218
fields. Overall, three cluster fields in very different directions on the 
sky contribute to the counts. Counts derived on a cluster by cluster
basis (Metcalfe et al.\,1999) were found to be consistent among the
clusters. 

However, it is not possible to exclude some impact of cosmic variance on
the faintest 15\,$\mu$m data point.  Yet the clearly
apparent internal consistency of the reported results (Figs.\ref{countslw3_diff}, \ref{countslw3_integ},
\ref{countslw2_diff} and \ref{countslw2_integ}), and the
excellent agreement with  models for the diffuse infrared
background (Figs.\ref{lari_mod}, \ref{IBL_fig}), underline the fact that no indications 
have been found that we are encountering any effects of cosmic variance.

\vskip -8pt
\subsubsection{Source count plots}
\label{sec:counts_plots}

\begin{figure*}
\centering
\includegraphics[angle=270,width=0.75\textwidth,bb=60 60 540 660]{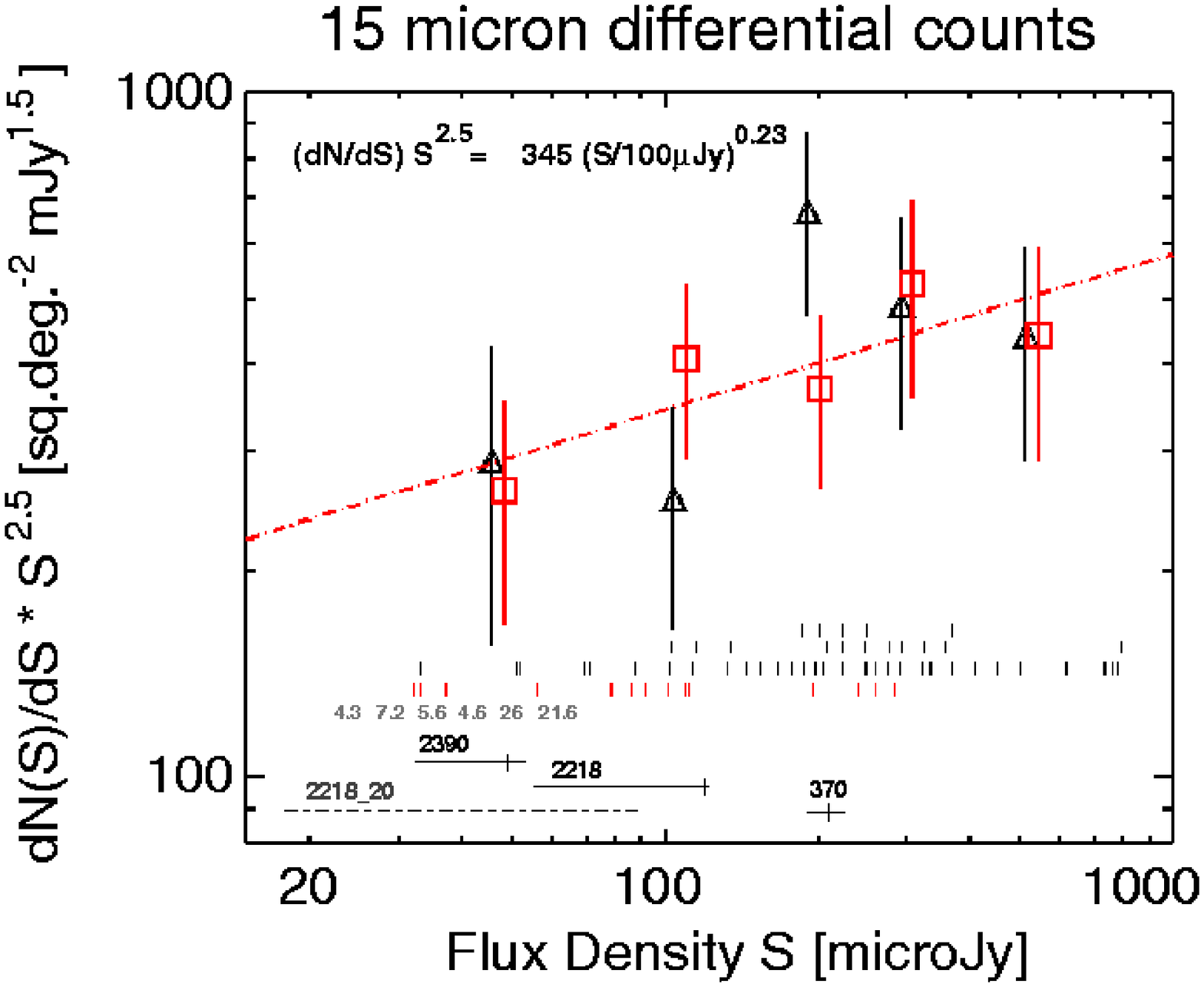}
\caption[]{Lensing-corrected 15\,$\mu$m differential source counts
accumulated from the observations of the three lensing clusters and normalised to the Euclidean 
distribution. The
triangles correspond to the lensing-and-completeness-corrected counts
to 80\% completeness. The squares represent the lensing-and-completeness-corrected 
counts to 50\% completeness. The broken line
through the data points is a simple fit to the 50\% completeness data.  The
horizontal rows of thin vertical bars indicate the flux-densities for
each of the sources used to make the counts, after correction for
lensing gain. Where sources have very similar fluxes the bars have
been offset vertically so that they can all be recognised. The bottom row indicates
sources counted between the 50\% and 80\% completeness threshold.  The three upper rows represent
sources counted above the 80\% completeness limit. The six numbers under the
vertical lines list the detection significance ($\times\sigma$) for the
six intrinsically weakest sources counted. (Not necessarily the
apparently weakest, because of lensing.) The horizontal bars near the
bottom of the plot show, for the faintest source counted in each cluster
field, the apparent brightness of the source (right end of bar) and the
intrinsic brightness after correction for lensing (left end). The detection limit for that cluster
in the absence of lensing is indicated in each case by the intersecting 
vertical bar. The dashed horizontal line labelled `2218\_20' indicates in a
similar manner the apparent and intrinsic brightness of the
intrinsically faintest 15\,$\mu$m source listed in the photometric tables 
for this survey (ISO\_A2218\_20).
}
\label{countslw3_diff}
\end{figure*}

\begin{figure*}
\centering
\includegraphics[angle=270,width=0.8\textwidth,bb=30 90 540 680]{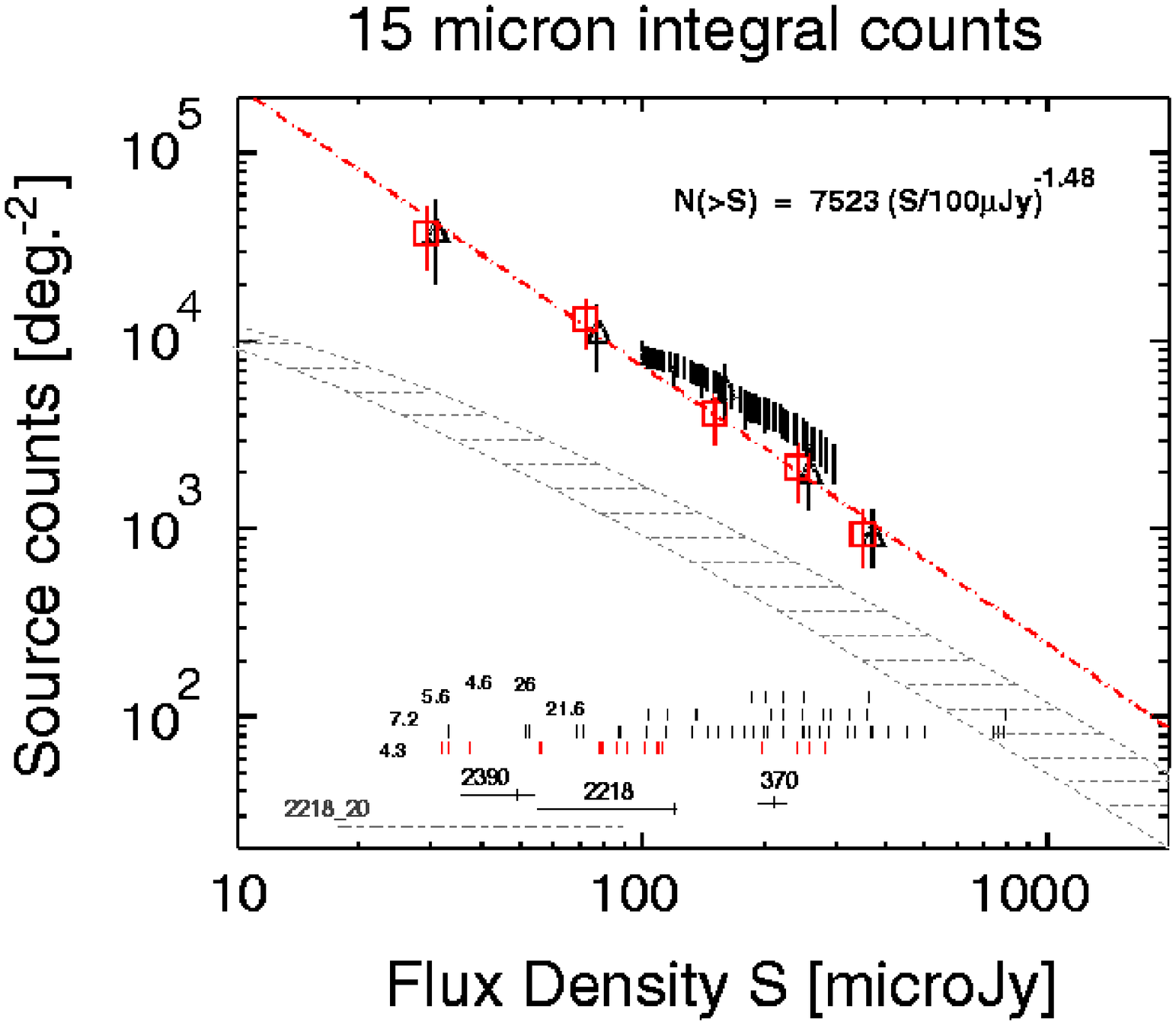}
\caption[]{Lensing-corrected 15\,$\mu$m integral source counts
accumulated from the observations of the three lensing clusters.  The
triangles correspond to the lensing-and-completeness-corrected counts
to 80\% completeness. The squares represent the lensing-and-completeness-corrected 
counts to 50\% completeness. The broken line
through the data points is a simple fit to the data. As explained in Gruppioni et al.\,(2002),
the hatched area represents the range of possible expectations from
no--evolution models normalised to the IRAS 12\,$\mu$m Local Luminosity
Function (upper limit from Rush et al.\,1993; lower limit from Fang et
al.\,1998). The black vertical hatching is the set of error bars for the HDFN counts of Elbaz et
al.\,(1999) \& Aussel et al.\,(1999a). Other features in the plot are as
described in the caption to Fig.~\ref{countslw3_diff}.
}
\label{countslw3_integ}
\end{figure*}

Figs.\ref{countslw3_diff} \& \ref{countslw3_integ} present respectively
lensing-corrected 15\,$\mu$m differential and integral source counts
accumulated from the observations of the three lensing clusters. 

\begin{figure*}
\centering
\includegraphics[angle=270,width=0.75\textwidth,bb=35 60 560 690]{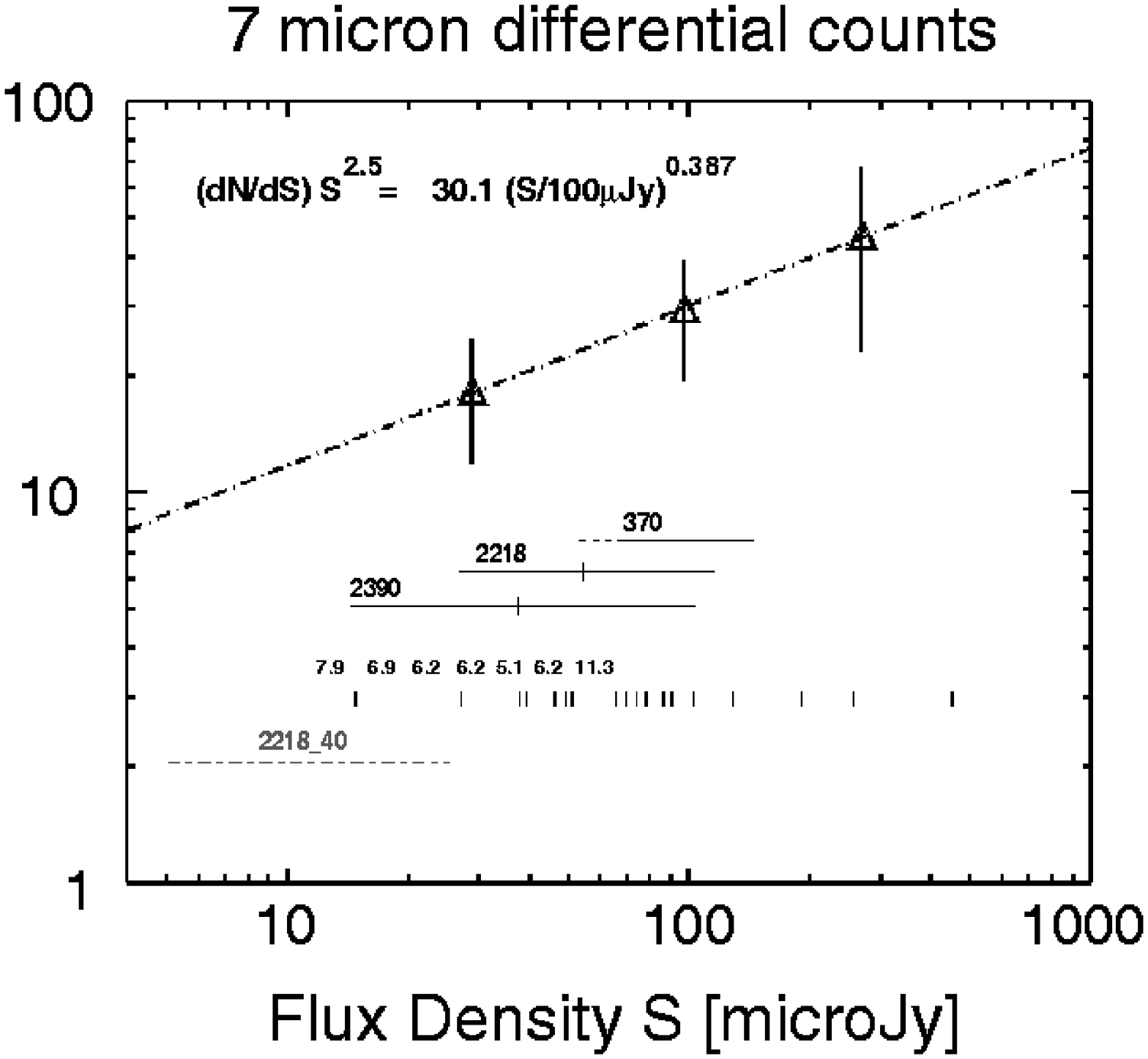}
\caption[]{ Lensing-corrected 7\,$\mu$m differential source counts
accumulated from the observations of the three lensing clusters and normalised to the Euclidean 
distribution.  The
triangles correspond to the lensing and completeness corrected counts
to 80\% completeness.  Other features in the plot are as
described in the caption to Fig.~\ref{countslw3_diff}, but the dashed
horizontal line labelled `2218\_40' indicates (in the manner described for
Fig.~\ref{countslw3_diff}) the apparent and intrinsic brightness of the
 intrinsically faintest 7\,$\mu$m source listed in the photometric tables for this survey
(ISO\_A2218\_40).
}
\label{countslw2_diff}
\end{figure*}

\begin{figure*}
\centering
\includegraphics[angle=270,width=0.75\textwidth,bb=40 40 560 680]{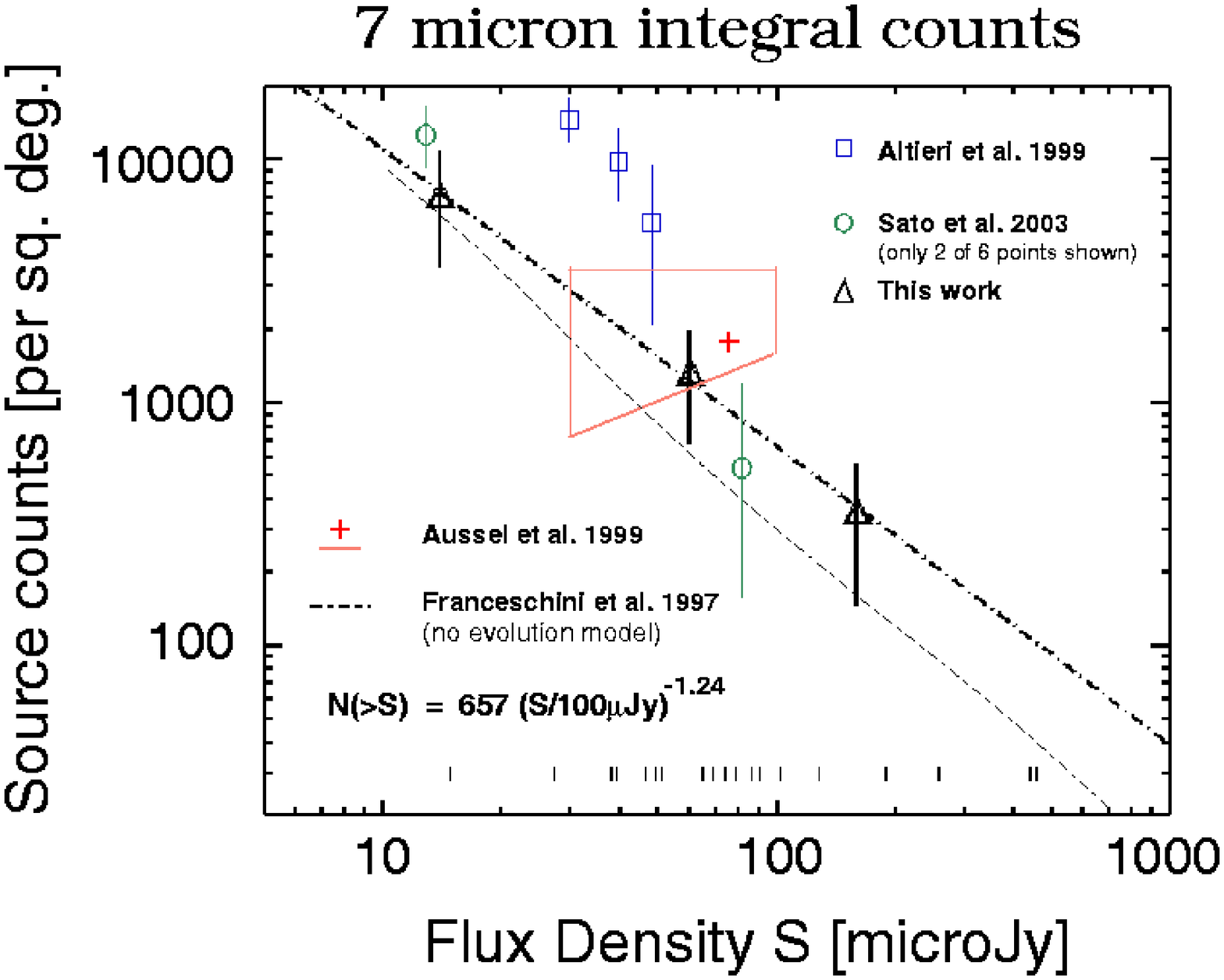}
\caption[]{ Lensing-corrected 7\,$\mu$m integral source counts
accumulated from the observations of the three lensing clusters. The
triangles correspond to the lensing and completeness corrected counts
to 80\% completeness.  The broken line through the data points is a
simple fit to the data. The low dashed line corresponds to the
no-evolution model of Franceschini et al.\,(1997).  The polyhedron containing the (+)
symbol indicates the 1-sigma error-box and best estimate 7\,$\mu$m
counts value of Aussel et al.\,(1999a), based on Hubble Deep Field
observations. The circles and squares are as indicated on the plot.
Other features in the plot are as described in the captions to
Fig.~\ref{countslw3_diff} and Fig.~\ref{countslw2_diff}.
}
\label{countslw2_integ}
\end{figure*}

\begin{figure*}
\centering
\includegraphics[angle=270,width=0.8\textwidth,bb=30 40 540 680]{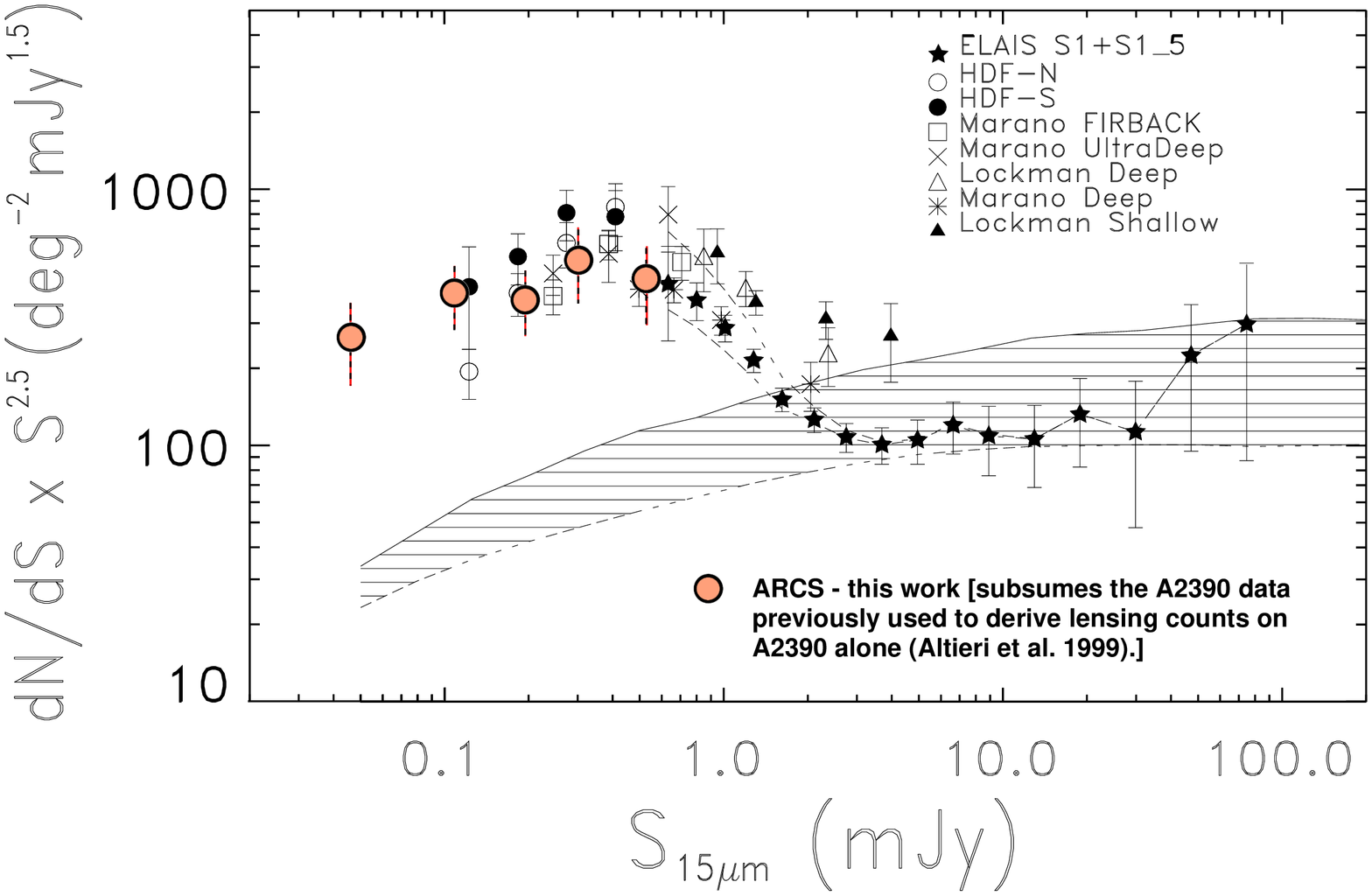}
\caption[]{Adapted from Gruppioni et al.\,(2002) figure 8.  Differential
source counts at 15\,$\mu$m normalised to the Euclidean distribution
($dN/dS \propto S^{-2.5}$). The results from various ISO surveys are
shown in the figure, including the present lensing survey, which
extends the plot to fainter sources.  As explained in Gruppioni et al.\,(2002),
the hatched area represents the range of possible expectations from
no--evolution models normalised to the IRAS 12\,$\mu$m Local Luminosity
Function (upper limit from Rush et al.\,(1993); lower limit from Fang et
al.\,(1998)).
}
\label{lari_mod}
\end{figure*}

\begin{figure*}
\centering \vspace*{+2cm}
\includegraphics[angle=0,width=0.75\textwidth,bb=40 40 560 680]{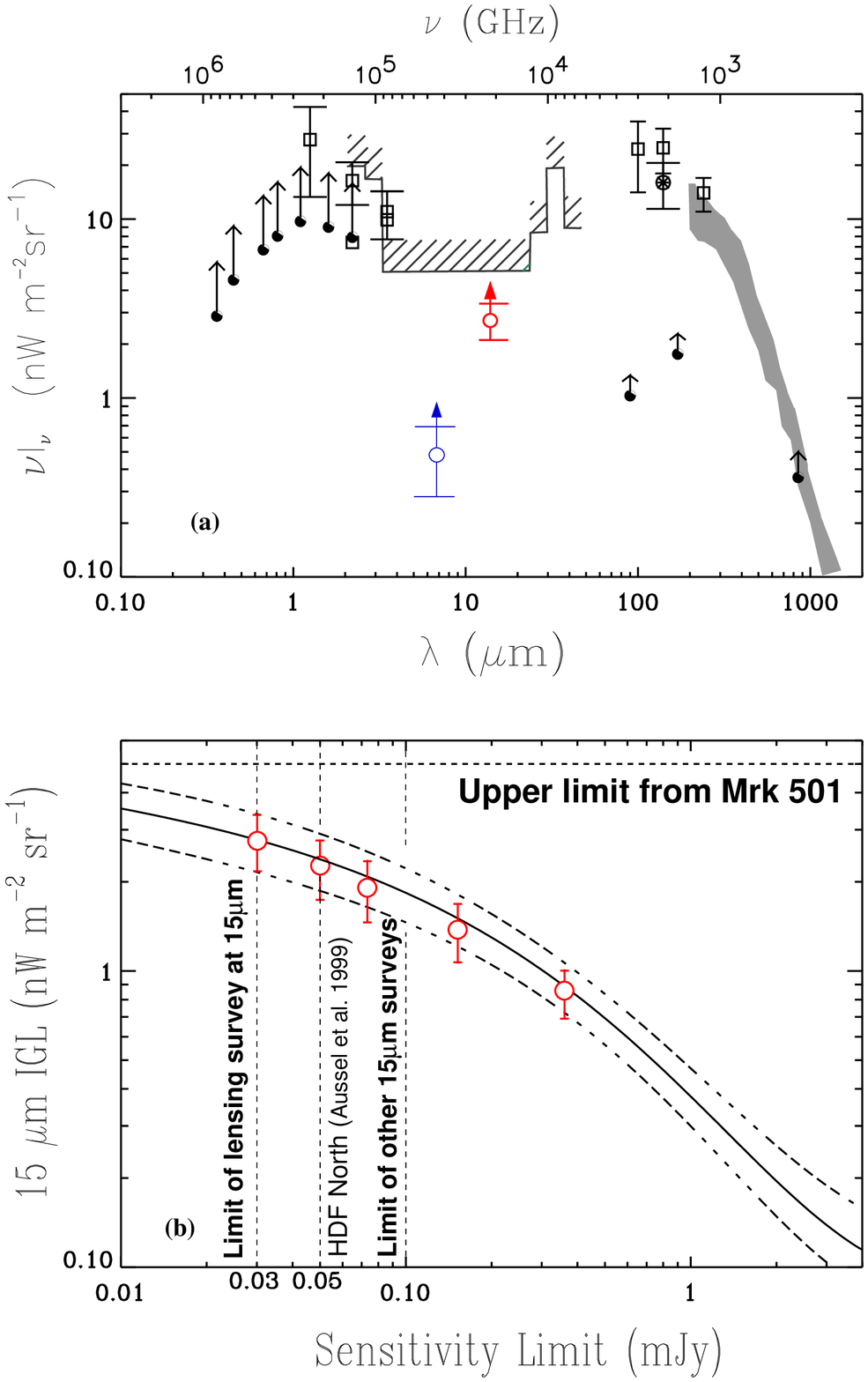}
\caption[]{ This figure is adapted from figs.3 and 10 of Elbaz et
al.\,(2002). (a) presents 
 limits to, and measurements of, the  Integrated Galaxy Light
(IGL, filled dots) and Extragalactic Background Light (EBL, open
squares, grey area) from the UV to sub-millimeter, as described in
the Elbaz et al.\,(2002) figure 10.  EBL measurements from COBE:
200-1500\,$\mu$m EBL from COBE-FIRAS (grey area, Lagache et al.
1999), 1.25, 2.2, 3.5, 100, 140\,$\mu$m EBL from COBE-DIRBE (open
squares). IGL in the U,B,V,I,J,H,K  bands from Madau \& Pozzetti
(2000). The upper end of the arrows indicate  the revised values
suggested by Bernstein et al. (2002, a factor of two  higher). The
hatched upper limit at 5 $\times 10^{-9}$ Wm$^{-2}$ sr$^{-1}$ is
obtained from the absorption of TeV photons with the extragalactic
infrared background. We have concatenated our 15\,$\mu$m counts with
higher flux data from the Elbaz et al. paper to arrive at a deeper
and so higher value of the 15\,$\mu$m IGL, 
(2.7$\pm$0.62)$\times$10$^{-9}$\,W\,m$^{-2}$\,sr$^{-1}$.  We
have replaced the Altieri et al.\,(1999) value for the 7\,$\mu$m
integral which Elbaz et al.\,(2002) plotted with the revised value
(0.49$\pm$0.2)$\times$10$^{-9}$\,W\,m$^{-2}$\,sr$^{-1}$. 

(b) shows values of the integral over the 15\,$\mu$m source counts evaluated to various
flux limits and plotted over the Elbaz et al.\,(2002) Fig.3 fit to the counts of other
ISOCAM deep surveys.  Our data points extend the 
flux-limit of the observations and remain consistent with the fit. The dashed lines
correspond to 1-$\sigma$ error bars obtained by Elbaz et al.\,(2002) fitting the
upper and lower limits of the source counts from other surveys.  The
error bars on our overplotted data points are obtained by integrating
over the upper and lower limits of the differential counts values.}
\label{IBL_fig}
\end{figure*}

{\noindent Figs.\ref{countslw2_diff} \& \ref{countslw2_integ} present respectively
lensing-corrected 7\,$\mu$m differential and integral source counts
accumulated from the observations of the three lensing clusters. The captions 
contain a detailed description of the figures.}

\begin{table}[!h]
\caption{\em Number of 15\,$\mu$m and 7\,$\mu$m non-stellar,
non-cluster, field sources falling above the 80\% and 50\% completeness
thresholds for the survey.
}
  \label{tab:numbers}
   \begin{center}
    \leavevmode
     \scriptsize 
      \begin{tabular}{ccccccccccc}
       \hline\hline
        \noalign{\smallskip}
    Filter  & Cluster &   no. of    &   no. of     \\
            &         &  sources    &  sources     \\
            &         & above  80\% & above  50\%  \\\hline
\noalign{\smallskip}             
 15\,$\mu$m &  A370   &     12      &      16      \\
\noalign{\smallskip}          
            &  A2218  &     23      &      29      \\
\noalign{\smallskip}         
            &  A2390  &     13      &      17      \\
\noalign{\smallskip}
\noalign{\smallskip}
  7\,$\mu$m &  A370   &      5      &     N/A      \\
\noalign{\smallskip}          
            &  A2218  &     11      &     N/A      \\
\noalign{\smallskip}         
            &  A2390  &      5      &     N/A      \\
\noalign{\smallskip}
   \hline
      \end{tabular}
   \end{center}
\end{table}

\begin{table}[!h]
\caption{\em Number of 15\,$\mu$m and 7\,$\mu$m countable sources per
             flux-density bin in the source-count plots
}
  \label{tab:binnumbers}
   \begin{center}
    \leavevmode
     \scriptsize 
      \begin{tabular}{cll}
       \hline\hline
        \noalign{\smallskip}
    Filter  &    no. of sources above              &    no. of sources above                 \\
            &   80\% compl. in bin                 &   50\% compl. in bin                    \\\hline
\noalign{\smallskip}             
            &  1 \ \ 2 \ \ \ 3 \ \ \ 4 \ \ \ 5 \ \ &  1 \ \ \ 2 \ \ \ 3 \ \ \ 4 \ \ \ 5 \ \  \\ 
\noalign{\smallskip}             
 15\,$\mu$m &  5 \ \,10 \ \ 13 \ \ 10 \ \ 10       &  9 \ \ 16 \ \ 15 \ \ 12 \ \ 10          \\
\noalign{\smallskip}
\noalign{\smallskip}
  7\,$\mu$m &  7 \ \ 8 \ \ \ 4  -- Two sources     & fell outside any used bin               \\
\noalign{\smallskip}
   \hline
      \end{tabular}
   \end{center}
\end{table}


\section{Discussion} 
\label{sec:discussion}

\subsection{Source counts and Log N-Log S distributions}
\label{sec:discuss_counts_plots}

Several independent field surveys were conducted in the mid-infrared using 
ISOCAM (e.g. Rowan-Robinson et al., 1997; Taniguchi et al., 1997; 
Altieri et al., 1999; Aussel et al., 1999a; Elbaz et al., 1999 \& 2000; 
Serjeant et al., 2000; Oliver et al., 2000; Lari et al., 2001; 
Gruppioni et al., 2002; Sato et al., 2003). 
Many of their results have been summarised in Elbaz et al.\,(1999 \& 2002). 
The Elbaz et al. (1999) paper includes all counts from the Lockman Hole, 
Marano Field, HDFN, HDFS and the results of our earlier work on A2390 
(Altieri et al. 1999).

Fig.~\ref{lari_mod}, adapted from figure 8 of Gruppioni et al.\,(2002),
compares the 15\,$\mu$m differential source counts from this work with
corresponding results from various ISO surveys, as identified in the
figure. The 15\,$\mu$m lensing corrected source counts from the three clusters 
presented here are in excellent agreement with the other surveys where the
flux ranges overlap. (The A2390 lensing points from Altieri et al. (1999) which 
appear in the figure in Lari et al. are here revised and subsumed in our global analysis 
of the three lensing clusters.)\\



Extending, and confirming, the results of the
earlier surveys, the  15\,$\mu$m counts in Fig.~\ref{countslw3_diff} \&
Fig.~\ref{countslw3_integ} show a steadily increasing excess by a
factor of order 10 over no-evolution  models (see references in figure
captions).  The slope of the 15\,$\mu$m differential counts is
-2.28 down to the 30\,$\mu$Jy limit. These  source
densities favour extreme evolution models. The mid-infrared-luminous galaxies
have been interpreted (Aussel et al.\,1999b; Elbaz et al.\,1999 \&
2000; Serjeant et al.\,2000) as a population  of dust-enshrouded star
forming galaxies. Within the current detection limits, they have a
median redshift of about 0.8 over all CAM deep surveys (Elbaz et al. 2002). 
The median z of all field (non-cluster) galaxies in the present work 
is 0.6, with a value of 0.65 applying to the background lensed sources.
 We attribute the difference between the Elbaz-quoted value and the value
from this work to two possible sources:
(a) spectroscopic  redshift values were not available for all counted 
sources. For
sources lacking spectroscopic  redshift values we inferred that they
are  non-cluster sources based on strong IR colour rules derived on
the  known redshift set (see Sec.\ref{sec:redshift}. It seems
reasonable to  suggest that the sources lacking spectroscopic
redshift values will be the optically  fainter, therefore more
distant, sources.  Now we determined our median value from the set of
sources for which we have spectroscopic z values.  Therefore our
median redshift estimate is likely to be biased towards lower
redshift. (b) As discussed in Sec.\ref{sec:cosmic_var}, we cannot
rule out  some impact of cosmic variance on our results, due to the
limited field areas surveyed.

\vskip 9pt
As pointed out in Elbaz et al. (2002), the 15\,$\mu$m number counts
cannot be explained without invoking either a strong evolution of 
the whole luminosity function (Xu 2000, Chary \& Elbaz 2001) or 
preferentially of a sub-population of starburst galaxies evolving both 
in luminos-
\newpage \noindent ity and density
(Franceschini et al. 2001, Chary \& Elbaz 2001, Xu et al. 2001). 

The evidence for strong evolution does not show up so clearly in the
7\,$\mu$m counts (Fig.~\ref{countslw2_diff} \&
Fig.~\ref{countslw2_integ}) which, within the error bars, are
consistent with the Franceschini et al.\,(1997) no-evolution model. 
This is not surprising, because the powerful starbursts associated with
strong evolution are hidden in the optical and near-infrared by 
absorbing clouds of dust. They become apparent through the reprocessing
of the starburst radiation into Unidentified Infrared Band (UIB) dust 
emission. (For a local example see Vigroux et al. 1996; Mirabel et al.
1998).  For the
redshift range of the  observed mid-infrared population, i.e. beyond
z=0.4, the 7\,$\mu$m  filter bandpass band is no longer sensitive to
the UIB emission, but is dominated by redshifted stellar light from the
rest-frame near-IR.  On the other hand, at similar redshifts the UIB
emission associated with dust absorbed starburst emission has been 
redshifted into the 15\,$\mu$m band, and falls in that band for
redshift  values up to $\sim$2.  

The redshift range of the 7\,$\mu$m known-z non-cluster field sources
is 0.026 to 2.8, with a median value of 0.55. 
The slope of the 7\,$\mu$m
differential counts is -2.11 down to the 14\,$\mu$Jy limit. It is 
important to note that, of the 20 sources used to derive the three
7\,$\mu$m data points in Figs.\ref{countslw2_diff} \& \ref{countslw2_integ}, 
fourteen have spectroscopic redshifts, one is
morphologically an obvious foreground elliptical  (ISO\_A370\_07+),
three (ISO\_A2218\_17, \_51, and \_67) have been securely identified as
galaxies based on SED modelling (Biviano, private communication), one 
(ISO\_A2218\_61a) has been classified
morphologically (as S0) from HST images by Smail et al.\,(2001), and  one
(ISO\_A2390\_22) has galactic, not stellar, mid-infrared colour.  So
our 7\,$\mu$m counts are not at all contaminated by stars, nor are they
contaminated by  cluster galaxies (see Sec.\ref{sec:redshift},
above).

\subsection{Extragalactic Infrared Background}
\label{sec:discuss_EIB}

The population of mid-infrared bright
galaxies observed by ISOCAM at 15\,$\mu$m can account for the bulk of
the Cosmic Infrared Background (CIRB), so that a large fraction of stars
seen today must have formed in dusty starbursts, such as those affecting
the ISOCAM galaxies (Elbaz et al.\, 2002, Lari et al, 2001, Altieri et
al. 1999). 
 It is interesting that Elbaz et al.\,(2002) also remark 
that the {\it unlensed} ISOCAM and SCUBA surveys do not sample the same
redshift and luminosity ranges, and hence show virtually no overlap
(Hughes et al.\ 1998), but that greater overlap of the samples is
expected in the deeper, lensed, cases.  Indeed we find that from a
sample of 15 SCUBA sources reported in the three clusters discussed in
this paper (Ivison et al.\,1998; Smail et al.\,2002; Cowie et al.\,2002) 
seven are seen in our ISOCAM 15\,$\mu$m sample. [I.e. sources
ISO\_A370\_09, ISO\_A370\_17, ISO\_A2390\_43, ISO\_A2390\_37 (cD galaxy), 
ISO\_A2390\_19, ISO\_A2390\_28a and ISO\_A2390\_42].  We will discuss
these cases in more detail in a later paper.

We concatenate the source counts obtained from the sample of lensed
background sources (and the few foreground sources) to the extensive
results of the other ISOCAM field surveys reported by Elbaz et
al.\,(2002).  Integrating in the  achieved flux density range from
30$\mu$Jy to 50 mJy, we resolve 
(2.7$\pm$0.62)$\times$10$^{-9}$\,W\,m$^{-2}$\,sr$^{-1}$ of the 
15\,$\mu$m infrared background into discrete sources. 
We similarly have  resolved a significant fraction of the 7\,$\mu$m
diffuse infrared  background, finding a lower value than reported in
Altieri et al. (1999), but agreeing well with the 7\,$\mu$m counts of
Sato et al. (2003) recorded on the SSA13
field.  The 7\,$\mu$m integral relies entirely on the lensing counts 
in the limited flux-density range from 14\,$\mu$Jy and 460\,$\mu$Jy to
resolve  (0.49$\pm$0.2)$\times$10$^{-9}$\,W\,m$^{-2}$\,sr$^{-1}$, or
$\sim$10\%, of the 7\,$\mu$m infrared background. In all cases the
error  limits are 1$\sigma$ values. 

We attribute the lower values found here for the IGL integrals, 
relative to the Altieri et al. (1999) value, to a much more thorough
rejection of cluster galaxy contamination (in the 7\,$\mu$m case),
and generally, to a much more rigourous Monte Carlo analysis of 
the completeness threshold performed in this work over the several 
intervening years.

Fig.~\ref{IBL_fig}(a) and (b) are adapted from figures 10 and 3,
respectively, of Elbaz et al.\,(2002).  (a) gives current limits to,
and  measurements of, the background radiation from UV to
sub-millimetre wavelengths, as described in Elbaz et al.\,(2002)
figure 10.   The two mid-infrared points take into account the results
of the work reported here, placed into the context of all the  ISOCAM
surveys.  The  15\,$\mu$m resolved Intergalactic Light (IGL) integral
benefits from the results of shallower surveys to reach deeper, and so
higher, integral values which are, as can be seen in
Fig.~\ref{IBL_fig}(b), consistent with the fit derived by Elbaz et
al.\,(2002), but extending to fainter flux-densities the experimental
confirmation of their fit.

 The spectra of high energy gamma-ray sources are predicted to be
modified by absorption effects caused by the pair-production
interaction of the gamma-rays with the extragalactic background light
(EBL) (Stecker, de Jager and Salamon 1992).  The peak absorption occurs
at wavelengths of $\lambda(\mu m)$ = 1.24 $E_{\gamma} (TeV)$ and
infrared photons are very effective at attenuating gamma-rays of
energies $E_{\gamma} \sim$ 1-100 TeV (see Hauser and Dwek 2001 for a
recent review).  The probing of the infrared background has recently
become practical with a range of gamma-ray observatories and
observations that have recently been reviewed by Catanese and Weekes
(2002).  The spectra of the high energy gamma rays from Mkn 421 and 501
have been used to place an upper limit of 5.0 $\times 10^{-9}$
Wm$^{-2}$ sr$^{-1}$ on the infrared background around 15 $\mu$m, as
shown by the hatched region in Fig 17(a) (Aharonian et al. 1999; de
Jager and Stecker 2002; Stanev and Franceschini 1997; Malkan and
Stecker 2001; Renault et al. 2001; Hauser and Dwek 2001).  The
contribution reported here to the infrared background light is 55\% of
the upper limit and most of the infrared background is now resolved
into discrete sources (Figs. 17(a) and (b)).

However the recent detections of TeV gamma rays (Horan et al.
2002: Petry et al. 2002: Aharonian et al. 2002) from the BL Lac
H1426+428 at z = 0.129, well beyond the redshifts of Mkn 421
(z=0.03) and Mkn 501 (z=0.034), imply severe absorption of TeV
photons by the optical and infrared backgrounds.  The value of the
15 $\mu$m background presented here has important consequences
for the TeV emission from H1426+428 because it implies a factor of
20 attenuation for the TeV photons, additional to the factor for
Mkn 501.  It is now reasonable to expect that further
observations of TeV gamma ray blazars will determine both the
optical and infrared extragalactic backgrounds (Aharonian et al.
2002; Weekes et al. 2002).

\subsection{Star formation history}
\label{sec:discuss_SF}

We take the opportunity to underline here the particular importance of
the ISOCAM mid-infrared observations in constraining important aspects of the
global history of star formation (Elbaz et al.\,2002).  The Madau plot
underestimates the star formation rate density per comoving volume in
the universe because in the first version (Madau et al.\,1996) no
correction for extinction was included, and in subsequent work (Madau
et al.\,1999) and other studies (e.g. Adelberger \& Steidel, 2000;
Meurer et al. 1999) the extinction correction is done assuming the
Calzetti law (Calzetti 1997) or the beta-slope technique from Meurer
et al.\,(1999). These techniques used to correct for extinction also
underestimate the true star formation rate which is measured in the mid
and far IR, as shown in the more recent paper by Goldader et
al.\,(2002) (see their Fig.1).

Luminous infrared galaxies appear to play a major role in the star
formation events that took place during the last 8-10 Gyr (since
z$\sim$1-1.2), as shown by the survey results already discussed and their
interpretation in terms of  contribution to the star formation history
and cosmic infrared background (Elbaz et al. 2002;  Chary \& Elbaz
2001). The initial Madau plot, from rest-frame UV data,
was missing two thirds of the star formation activity in galaxies, and
correcting for extinction with the UV-slope technique, as done by
several authors, still misses most of it, since this law does not work
for luminous IR galaxies (see Goldader et al. 2002, Fig.1).

\section{Conclusions}  
\label{sec:conclude}

In this paper, we have described the use of the ISOCAM camera on board
ESA's Infrared Space Observatory mission to employ clusters of galaxies
as natural lenses to extend the mid-Infrared source counts of field
sources to the deepest levels, at both 7\,$\mu$m and 15\,$\mu$m.  A
flux selected subset of the sources above the 80\% and 50\%
completeness limits was used to derive cumulative source counts to a
lensing corrected sensitivity level of 30\,$\mu$Jy at 15\,$\mu$m, and
14\,$\mu$Jy at 7\,$\mu$m.  For the 7 $\mu$m bandpass cluster sources were 
numerous, but cluster contamination could be eliminated by reference to 
source spectroscopic redshift, and in a few cases by the strong correlation
between mid-IR colour and cluster membership.  The 15 $\mu$m source
counts, corrected for the effects of contamination by cluster sources
and lensing, confirm and extend earlier findings of an excess by a factor
 of 10 with respect to source models with no evolution.  

We resolve (2.7$\pm$0.62)$\times$10$^{-9}$\,W\,m$^{-2}$\,sr$^{-1}$ of
the 15\,$\mu$m infrared background into discrete sources by integrating
over our source counts, concatenated to those from other extensive ISOCAM 
surveys, in the flux range from 30$\mu$Jy to 50 mJy.  We resolve 
(0.49$\pm$0.2)$\times$10$^{-9}$\,W\,m$^{-2}$\,sr$^{-1}$ of the
7\,$\mu$m  infrared background into discrete sources by integrating
over our source counts in the flux range from 14\,$\mu$Jy to
460\,$\mu$Jy.  

Recent detections of TeV gamma rays from more distant BL Lac sources
serve to underline the opacity, to TeV photons, of the mid-infrared photon
densities reported in this work and give reason to suspect that the mid-infrared
flux-densities found may indeed account for a larger fraction of the diffuse
background than the 55\% mentioned just above in the discussion.

These results suggest that abundant star formation occurs in very
dusty  environments at z$\sim$1. The next Mid-Infrared (SIRTF, NGST)
and Far-Infrared (Herschel) mission, will clearly help in understanding
star-formation processes at high redshift.

We have demonstrated over an unprecedentedly large source sample that
gravitational lens models for galaxy clusters can be successfully used
to perform statistical source-counting surveys to depths otherwise
unachievable with the same instrumentation. We have presented a
photometric catalogue of 145 mid-infrared sources to unprecedented
depth, containing roughly 15\% of all deep survey sources detected in
the several extensive ISOCAM surveys.

When deep cosmological surveys are planned for future missions, the use of
nature's own telescopes to increase the depth of the surveys should not be
considered as a novelty. It should form an indispensible part of the core 
survey strategy from the outset.  The potential of this technique has only 
been illustrated in this work, which has drawn upon what has probably
been close to the minimum viable dataset for these purposes.  The more
clusters used in a lensing survey the more area is accumulated at the
highest amplification values.  As the plots in Fig.~\ref{how_deep1}
illustrate, lensing has the potential to make sources available at
intrinsic fluxes arbitarily fainter than the apparent flux limit for
deep blank--field surveys made with the same instrumentation.

\begin{acknowledgements} 
{\scriptsize
We acknowledge helpful discussions with Andrew Blain, Rob Ivison, Roser
Pell\`o, Ian Smail, Genevi\`eve Soucail and invaluable advice from
Laurent Vigroux.  We thank Sibylle Peschke for careful assistance with
the layout of the paper. We thank the referee, Richard Barvainis, for
encouraging and helpful comments and suggestions. JPK thanks CNRS for 
support. The ISOCAM data presented in this paper was analysed using
``CIA", a joint development by the ESA Astrophysics Division and the
ISOCAM Consortium. The ISOCAM Consortium is led by the ISOCAM PI, C.
Cesarsky.  This research made use of the NASA/IPAC Extragalactic
Database (NED) which is operated by the Jet Propulsion Laboratory,
California Institute of Technology, under contract with the National
Aeronautics and Space Administration.
}
\end{acknowledgements}

\end{document}